\documentclass[a4paper,aps,showpacs,floatfix,twocolumn,%
superscriptaddress]{revtex4}

\usepackage{graphicx}
\usepackage{amsmath}
\usepackage{amssymb}
\usepackage{gensymb}
\usepackage{color}

\begin{document}

\title{Studying Flow Close to an Interface by Total Internal Reflection \\
Fluorescence Cross Correlation Spectroscopy: \\
Quantitative Data Analysis
}

\author{R. Schmitz}
\affiliation{Max Planck Institute for Polymer Research,
Ackermannweg 10, 55128 Mainz, Germany}
\author{S. Yordanov}
\affiliation{Max Planck Institute for Polymer Research,
Ackermannweg 10, 55128 Mainz, Germany}
\author{H.~J. Butt}
\affiliation{Max Planck Institute for Polymer Research,
Ackermannweg 10, 55128 Mainz, Germany}
\author{K. Koynov}
\affiliation{Max Planck Institute for Polymer Research,
Ackermannweg 10, 55128 Mainz, Germany}
\author{B. D\"unweg}
\affiliation{Max Planck Institute for Polymer Research,
Ackermannweg 10, 55128 Mainz, Germany}
\affiliation{Department of Chemical Engineering, Monash University,
Clayton, Victoria 3800, Australia}

\date{\today}

\begin{abstract}

  Total Internal Reflection Fluorescence Cross Correlation
  Spectroscopy (TIR-FCCS) has recently (S. Yordanov et al., Optics
  Express 17, 21149 (2009)) been established as an experimental method
  to probe hydrodynamic flows near surfaces, on length scales of tens
  of nanometers. Its main advantage is that fluorescence only occurs
  for tracer particles close to the surface, thus resulting in high
  sensitivity. However, the measured correlation functions
  only provide rather indirect information about the flow parameters
  of interest, such as the shear rate and the slip length. In the present
  paper, we show how to combine detailed and fairly realistic theoretical
  modeling of the phenomena by Brownian Dynamics simulations with accurate
  measurements of the correlation functions, in order to establish
  a quantitative method to retrieve the flow properties from the
  experiments. Firstly, Brownian Dynamics is used to sample highly accurate
  correlation functions for a fixed set of model parameters. Secondly,
  these parameters are varied systematically by means of an
  importance-sampling Monte Carlo procedure in order to fit the
  experiments. This provides the optimum parameter values together
  with their statistical error bars. The approach is well suited
  for massively parallel computers, which allows us to do the data
  analysis within moderate computing times. The method is applied to
  flow near a hydrophilic surface, where the slip length is observed
  to be smaller than $10nm$, and, within the limitations of the
  experiments and the model, indistinguishable from zero.

\end{abstract}


\maketitle


\section{Introduction}
\label{sec:intro}

A good understanding of liquid flow in confined geometries is not only
of fundamental interest, but also important for a number of
industrial and technological processes, such as flow in porous media,
electro-osmotic flow, particle aggregation or sedimentation, extrusion
and lubrication. It is also essential for the design of micro- and
nano-fluidic devices, e.~g. in lab-on-a-chip applications. However, in
all these cases, an accurate quantitative description is only possible
if the flow at the interface between the liquid and the solid is
thoroughly understood \cite{Tabeling06, Vinogradova1999, Ellis04, Net05, %
Lauga07, Bonaccurso02, Neto03, Rod10, Guriyanova08, Feuillebois09}.
While for many years the so-called no-slip boundary
condition (relative velocity at the interface equal to zero) had been
successfully applied to describe macroscopic flows, more recent
investigations revealed that this condition is insufficient to
describe the physics when flows through channels with micro- and
nano-sizes are considered \cite{Net05, Lauga07}. On such small scales,
the relative contribution from a residual slip between liquid and solid
becomes important. This is commonly described by the so-called slip
boundary condition, which is characterized by a non-vanishing slip
length $l_s$, defined as the ratio of the liquid dynamic viscosity and
the friction coefficient per unit area at the surface. An equivalent
definition is obtained by taking the ratio of the finite surface flow
velocity, the so-called slip velocity $v_s$, and the shear rate at the
surface: $l_s = v_s/(dv/dz)_{z=0}$, where $z$ is the spatial direction
perpendicular to the surface, located at $z = 0$. This boundary
condition is the most general one that is possible within the
framework of standard hydrodynamics \cite{einzel1990boundary}; the
no-slip condition is simply the special case $l_s = 0$.

The experimental determination of the slip length, however, is
challenging, since high resolution techniques are needed to gain
sufficiently accurate information close to the interface. Hence, the
existence and the magnitude of slip in real physical systems, as well
as its possible dependence on the surface properties, are highly
debated in the community, and no consensus has been reached so
far. Clearly, a resolution of these controversies requires further
improvement of the experimental techniques.

To date, two major types of experimental methods, often called direct
and indirect, have been applied to study boundary slip phenomena. In
the indirect approach, an atomic force microscope or a surface force
apparatus is used to record the hydrodynamic drainage force necessary
to push a micron-sized colloidal particle versus a flat surface as a
function of their separation \cite{Ducker1991, Butt1991}. The
separation can be measured with sub-nanometric resolution, and the
force with a resolution in the $pN$ range. A high force is necessary to
squeeze the liquid out of the gap if the mobility of the liquid is
small. Conversely, if the liquid close to the surface can easily slip
on it, then a small force is necessary. From this empirical
observation a quantitative value of the slip length can be deduced
using an appropriate theoretical model \cite{Vinogradova1999,
  Vinogradova1995, Bonaccurso02}. While this approach is extremely
accurate at the nanoscale, it does not measure the flow profile
directly.

Direct experimental approaches to flow profiling in microchannels are
commonly based on various optical methods to monitor fluorescent
tracers moving with the liquid. Basically they can be divided into two
sub-categories.

The imaging-based methods use high-resolution optical microscopes and
sensitive cameras to track the movement of individual tracer particles
via a series of images \cite{Pit2000, Tretheway02, Jos05, Huang06, %
Lasne08, Bouzigues08, Li10}. While providing a real ``picture'' of
the flow, the imaging methods have also some disadvantages related
mainly to the limited speed and sensitivity of the cameras: relatively
big tracers are needed, the statistics is rather poor, and large
tracer velocities cannot be easily measured.

In the Fluorescence Correlation Spectroscopy (FCS) based methods the
fluctuations of the fluorescent light emitted by tracers passing
through a small observation volume (typically the focus of a
confocal microscope) are measured \cite{Rigler01}. Using correlation
analysis and an appropriate mathematical model the tracers' diffusion
coefficient and flow velocity can be evaluated \cite{Magde1978,
  Orden1998, Kohler2000, Gosch2000}. In particular, the so-called
Double-Focus Fluorescence Cross-Correlation Spectroscopy (DF-FCCS) that
employs two observation volumes (laterally shifted in flow direction)
is a powerful tool for flow profiling in microchannels
\cite{Bri99, Dit02, Lum03, Vin09}.  Due to the high sensitivity
and speed of the used photo detectors (typically avalanche
photodiodes) in the FCS based methods even single molecules can be
used as tracers. Furthermore, the evaluation of the velocity is based
on large statistics and high tracer velocities can be measured.

During the last two decades both the imaging and the FCS methods were
well developed to the current state that allows fast and accurate
measurements of flow velocity profiles in microchannels. The
situation, however, is different when the issue of boundary slip is
considered. Due to the limited optical resolution imposed by the
diffraction limit, it is commonly believed that these methods are less
accurate than the force methods discussed above and cannot detect a
slip length in the tens of nanometers range. On the other hand, the
possibility to directly visualize the flow makes the optical methods
still attractive and thus continuous efforts were undertaken to
improve their resolution. One of the most successful approaches in
this endeavor is Total Internal Reflection Microscopy (TIRM)
\cite{Axelrod1984}, which uses total internal reflection at the
interface between two media with different refractive indices, like,
e.~g., glass and water. This creates an evanescent wave that extends
into the liquid in a tunable region of less than $\sim 200 nm$ from
the interface. Optical excitation of the fluorescent tracers is then
possible only within this narrow region. During the last few years
TIRM was successfully applied to improve the resolution of particle
imaging velocimetry close to liquid-solid interfaces \cite{Huang06, %
  Lasne08, Bouzigues08, Li10}, and slip lengths in the order of tens
of nanometers were evaluated. With respect to FCS, however, TIR
illumination had, until recently, been limited to diffusion studies
only \cite{HasslerSecond05, Rie08}, while there were no reports on
TIR-FCS based velocimetry and slip length measurements.

With this in mind, we have recently developed a new experimental setup
that combines for the first time TIR illumination with DF-FCCS for
monitoring a liquid flow in the close proximity of a solid
surface \cite{Koy09}. Such a combination offers high normal
resolution, extreme sensitivity (down to single molecules), good
statistics within relatively short measurement times and the
possibility to study fast flows. Our preliminary studies have
shown, however, that the accurate quantitative evaluation of the
experimental data obtained with this TIR-FCCS setup is not
straightforward because the model functions needed to fit the auto-
and cross-correlation curves (and extract the flow velocity profile)
are not readily available. The standard analytical procedure to derive
these functions is \cite{Bri99, Dit02, Lum03}: (i) solve the
convection-diffusion equation with respect to the concentration
correlation function, (ii) insert the derived solution in the
corresponding correlation integral and (iii) solve it to finally get
the explicit form of the correlation functions. This procedure was
successfully used by Brinkmeier et al. \cite{Bri99} to derive
analytical expressions for the auto- and cross-correlation functions
obtained with double focus confocal FCCS (i.~e. with focused laser
beam illumination as opposed to the evanescent wave illumination in
our case), where it was assumed that the flow velocity and tracer
concentration are spatially constant, which simplifies the calculation
substantially. Such an assumption is reasonable if the observation
volumes (laser foci) are far away from the channel walls, in the same
distance. In the case of TIR-FCCS, however, the situation is different:
The experiments are performed in the proximity of the channel
wall and the distribution of the flow velocity inside the observation
volume has to be considered. Furthermore, the concentration of tracers
may also depend on $z$ due to electrostatic repulsion or hydrodynamic
effects. Finally the presence of a boundary, which must also be taken
into account in the theoretical treatment, further complicates the
problem. Therefore, a faithful description of the physics of TIR-FCCS
makes the problem of calculating the correlation functions (rather
likely) unsolvable in terms of closed analytical expressions.

For this reason, we rather resort to numerical methods, and in the
present paper describe and test the procedure that we have developed:
We employ Brownian Dynamics techniques to simulate the tracers' motion
through the observation volumes and generate ``numerical'' auto- and
cross-correlation curves that are consequently used to fit the
corresponding experimental data. This fitting is done via Monte Carlo
importance sampling in parameter space. The method is therefore fully
quantitative, while not being hampered by any difficulties in doing
analytical calculations. It should be noted that this approach also
provides a substantial amount of flexibility: The details of the
physical model are all encoded in the Brownian Dynamics simulation
which specifies how the tracer particles move within the flow. In the
present work we have assumed a simple Couette flow with a finite slip
length, while the particles are described as simple hard spheres with
no rotational degree of freedom, and no interaction with the wall
except impenetrability. It is fairly straightforward to improve on
these limitations, by, e.~g., including hydrodynamic and electrostatic
interactions with the wall, rotational motion of the spheres, or
polydispersity in the particle size distribution. Moreover, the
geometry of the observation volumes can be easily changed as well, and
we have made use of this possibility in our present work, but only to
some extent. Further refinements are left for future work, in which
the basic methodology would however remain unchanged.

To test the accuracy of the newly developed TIR-FCCS experimental
setup and the numerical data evaluation procedure, we have studied
aqueous flow near a smooth hydrophilic surface and evaluated the slip
length to be between $0$ and $10 nm$ (however with a systematic error
that is hard to quantify, and whose elimination would need a more
sophisticated theoretical model). It is commonly accepted
\cite{Lasne08, Bouzigues08, Li10, Idol1986, Jos05, Charlaix05, %
  Joly06, Honig07} that the boundary slip should be zero (or very
small) in this situation. Thus, our results indicate that TIR-FCCS
offers unprecedented accuracy in the $10nm$ range for the measurement
of slip lengths by an optical method. We believe that our result for
the slip length will be fairly robust, even if the physical model is
refined further. 

Section \ref{sec:exp} outlines the experimental setup, while Sec.
\ref{sec:exp_corrfunc} presents the experimental results and the
numerical fits. We find that the measured cross-correlation functions
deviate considerably from the model functions at short times, probably
as a result of some optical effects which at present we do not fully
understand.  However, we show a practical way to eliminate such
effects to a large extent, by means of a simple subtraction
scheme. The following parts then outline in detail how the theoretical
curves have been obtained: Firstly, Sec. \ref{sec:corrfunc} elucidates
the relation between the measured correlation functions and the
underlying dynamics of the tracer particles. We then proceed to
describe the Brownian Dynamics algorithm to sample the model
correlation functions (Sec. \ref{Sec: Sampling Algorithm}). Section
\ref{sec:subtraction} then provides a detailed theoretical analysis of
our subtraction scheme. In Sec. \ref{sec: stat_dat} we describe the
Monte Carlo method to find optimized parameter values of our
model. Section \ref{sec:results} then discusses our results, in
particular concerning the slip length; this is followed by a brief
summary of our conclusions (Sec. \ref{sec:conclus}).

\section{Experimental Setup}
\label{sec:exp}

\begin{figure}[t]
  \noindent
  \begin{centering}\hspace{-0.4cm}
     \includegraphics[keepaspectratio,width=0.48\textwidth]
     {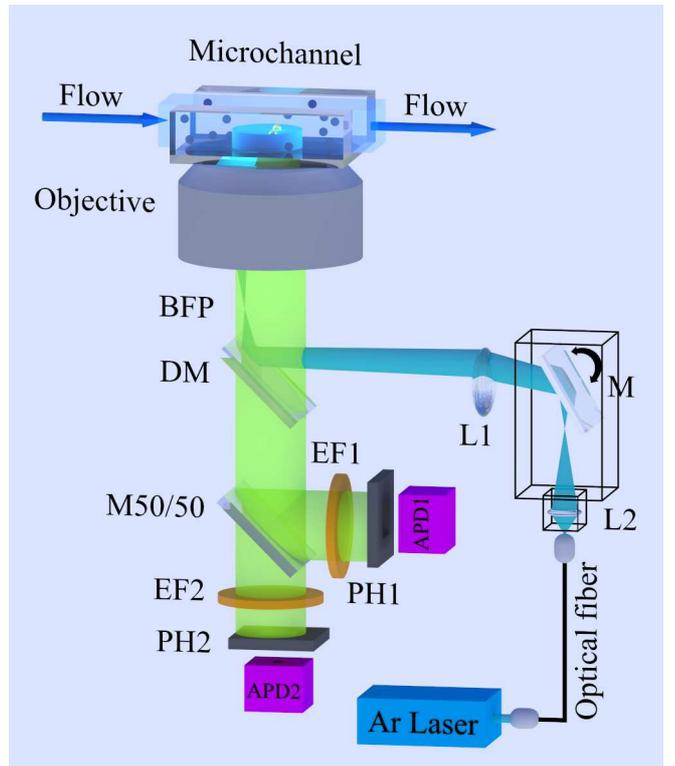}
  \end{centering}
  \caption{(Color) Scheme of the experimental TIR-FCCS setup. BFP -
    back focal plane of the objective; DM - dichroic mirror; M50/50 -
    neutral $50\%$ beam splitter; EF1, EF2 - emission filters; PH1,
    PH2 - pinholes; APD1, APD2 - avalanche photodiodes; L1 - tube
    lens; L2 - collimator lens; M - collimator's prism based
    mirror. Note that the two spatially separated observation volumes
    are created by shifting the pinholes PH1/PH2 in the
    $x$-$y$-plane. The cyan color indicates the excitation wavelength
    and the yellow-green color the fluorescence light, respectively.}
    \label{fig: TIR-FCCS setup}
\end{figure}
  
\begin{figure}[t]
  \noindent
  \begin{centering}\hspace{-0.4cm}
     \includegraphics[keepaspectratio,width=0.48\textwidth]
     {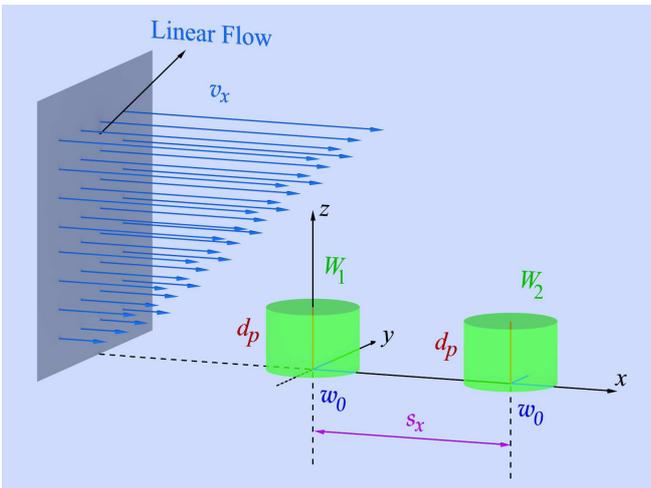}
  \end{centering}
  \caption{(Color) The coordinate system and the linear flow field
    employed in the TIR-FCCS experiment. $W_1$ and $W_2$ denote the
    shape and location of the observation volumes as seen by pinhole
    PH1 and pinhole PH2, respectively; $d_p$ is the penetration depth
    which defines the axial extent of the observation volume; $w_0$ is
    the typical extension of the observation volumes in the
    $x$-$y$-plane; $s_x$ indicates the observation volumes separation,
    center-to-center distance; $v_x$ is the velocity field in positive
    $x$ direction, which depends linearly on $z$.}
    \label{fig: observation volume plus flow}
\end{figure}

Since the TIR-FCCS experimental setup has already been described in
great detail elsewhere \cite{Koy09}, only a brief qualitative overview
of the basic ideas and quantities is given below. A scheme of the
experimental setup is shown in Fig. \ref{fig: TIR-FCCS setup}. It is
based on a commercial device (Carl Zeiss, Jena, Germany) that consists
of the FCS module ConfoCor2 and an inverted microscope Axiovert
200. The TIR illumination is achieved by focusing the excitation laser
beam ($488 nm$, Ar$^{+}$ Laser) on the periphery of the back focal
plane (BFP) of an oil immersion microscope objective with numerical
aperture $NA = 1.46$. This leads to a parallel laser beam which
emerges out of the objective and then enters the rectangular flow
channel through its bottom wall (Fig. \ref{fig: TIR-FCCS setup}). By
adjusting the angle of incidence above the critical angle ($\approx 61
\degree$ for the glass-water interface) total internal reflection is
achieved. In this situation only an evanescent wave extends into the
liquid and can excite the fluorescent tracers suspended in it. The
intensity distribution of this wave in the $x$-$y$-plane (parallel to
the interface) is Gaussian with a diameter of $\sim 30\mu m$ (at
$e^{-1}$). In the $z$ direction the intensity decays exponentially, $I
(z) = I_{0} \exp(-z/d_p)$. The characteristic decay length $d_p$, also
called penetration depth, depends on the laser wavelength $\lambda$,
the refraction indices of both media ($n_1$ - glass, $n_2$ - water)
and can be varied in the range $80-200 nm$ by changing the angle of
incidence. Thus the evanescent wave can excite only the tracers
flowing in the proximity of the channel wall. The produced
fluorescence light is collected by the same microscope objective and
is equally divided by passing through a neutral $50\%$ beam splitter
to enter two independent detection channels. In each channel the
fluorescent light passes through an emission filter and a confocal
pinhole to finally reach the detectors, two single photon counting
avalanche photodiodes (APD1, APD2). The pinholes PH1 and PH2 define
two observation volumes that are laterally shifted with respect to
each other along the flow direction as schematically shown in
Fig. \ref{fig: observation volume plus flow}. The center-to-center
distance $s_x$ between the two observation volumes can be continuously
tuned from $0$ to $3\mu m$. The signals from both channels are
recorded and correlated to finally yield the auto- and
cross-correlation curves that contain the entire information about the
flow properties, slip length and shear rate, close to the interface.

The experiments were performed with a rectangular microchannel of $L_y
= 4 mm$ width, $L_z = 100\mu m$ height and $L_x = 50 mm$ length
fabricated using a three-layer sandwich construction as described in
earlier work \cite{Koy09, Lum03}. The bottom channel wall at which the
TIR-FCCS experiments were performed was a microscope cover slide made
of borosilicate glass with a thickness of $170\mu m$, cleaned with
$2\%$ aqueous solution of Hellmanex and Argon plasma. The
root-mean-square roughness of the glass surface was in the range of
$0.3 nm$ and the water advancing contact angle below $5 \degree$
(hydrophilic surface). The flow was induced by a hydrostatic pressure
gradient, created by two beakers of different heights, where the water
level difference was kept constant by a pump. This allowed us to vary
the shear rate near the wall in the range $0 - 5000 s^{-1}$.

Carboxylate-modified quantum dots (Qdot585 ITK Carboxyl, Molecular
Probes, Inc.), with a hydrodynamic radius $R_H = 6.87 nm$, were used
as fluorescent tracers. The particles were suspended in an aqueous
solution of potassium phosphate ($K_2HPO_4$) buffer ($pH \simeq 8.0$,
concentration $6 mM$). The concentration of the quantum dots
was found from our data analysis (see below) as $\sim 30 nM$,
corresponding to roughly $18$ particles per $(\mu m)^3$.

\section{Correlation Curves}
\label{sec:exp_corrfunc}

The motion of the fluorescence tracers results in two time-resolved
fluorescence intensities $I_1(t)$ and $I_2(t)$, which were measured
with the two photo detectors. For the present system, we may safely
assume that it is ergodic and strictly stationary on the time scale of
the experiment, such that only time differences matter
\cite{Lakowicz06}. Therefore, we may define the intensity
fluctuations via
\begin{equation}
\delta I_i(t) = I_i (t) - \langle I_i \rangle ,
\end{equation}
where $\langle \cdot \rangle$ denotes a time average or, equivalently,
an ensemble average, and evaluate the time-dependent auto- and
cross-correlation functions via the definition
\begin{equation}
  \label{eq: G correlation function matrix}
  G_{ij}(t) =
  \frac{\langle \delta I_i(0) \delta I_j(t) \rangle}
  {\langle I_i \rangle \langle I_j \rangle} .
\end{equation}
It should be noted that possible small differences in the sensitivity
of the photo detectors or in the illumination of the pinholes cancel
out, since in Eq. \ref{eq: G correlation function matrix} only
\emph{ratios} of intensities occur. $G_{11}$ and $G_{22}$ are the two
autocorrelation functions of pinholes $1$ and $2$, respectively, while
$G_{12}$ and $G_{21}$ are the forward and backward cross-correlation
functions, respectively. It should be noted that in the presence of
flow $G_{12}$ and $G_{21}$ differ substantially. In the limit of the
two pinholes being located at the same position, the intensities $I_1$
and $I_2$ coincide, such that in this case all four entries of the
matrix $G_{ij}$ are identical.

\begin{figure*}[p]

\begin{tabular}{c c}
\parbox{0.05\textwidth}{(a)}
\parbox{0.35\textwidth}{
  \includegraphics[keepaspectratio,width=0.32\textwidth]
  {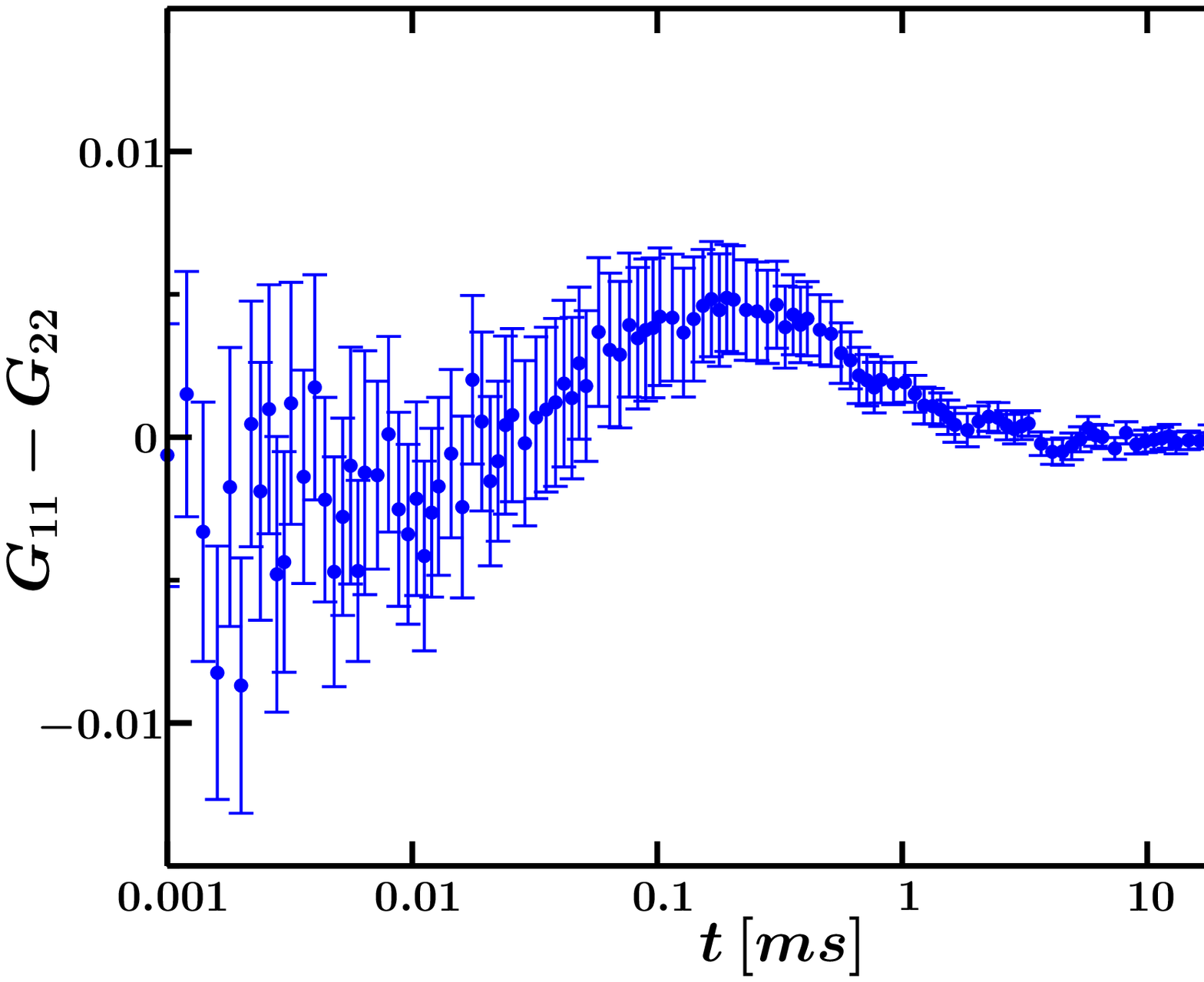} }
&
\parbox{0.05\textwidth}{(b)}
\parbox{0.35\textwidth}{
  \includegraphics[keepaspectratio,width=0.32\textwidth]
  {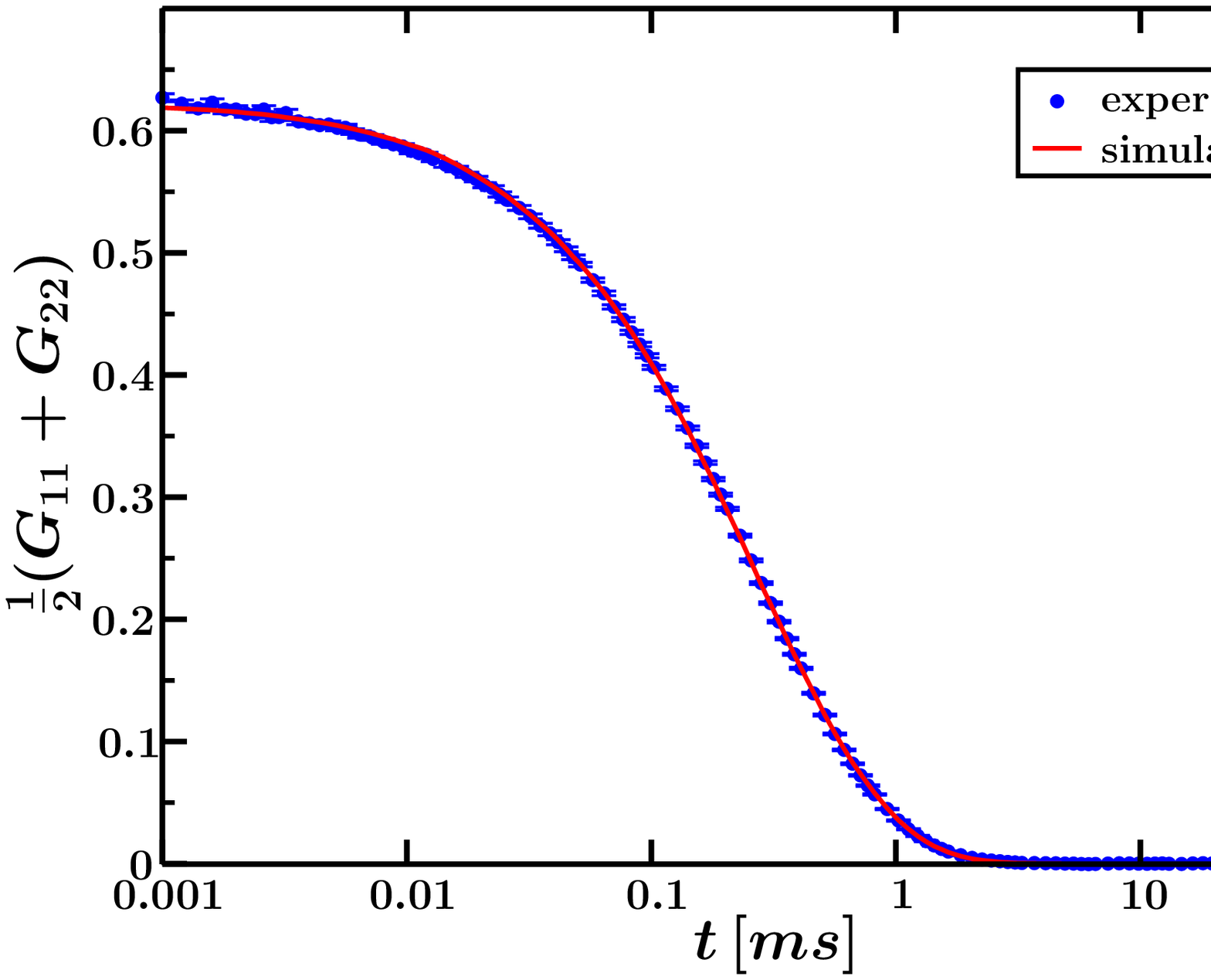} }
\\
\parbox{0.05\textwidth}{(c)}
\parbox{0.35\textwidth}{
  \includegraphics[keepaspectratio,width=0.32\textwidth]
  {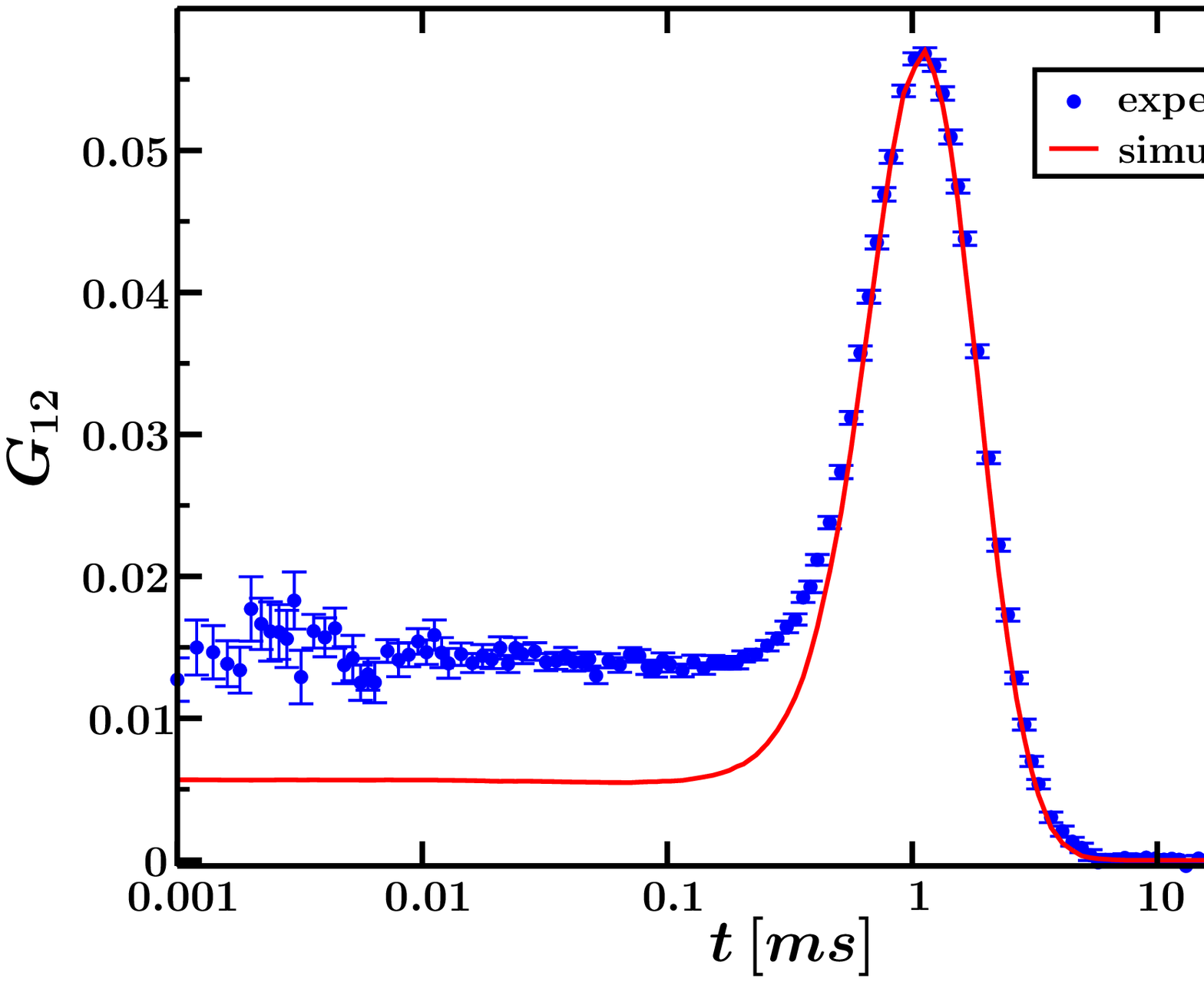} }
&
\parbox{0.05\textwidth}{(d)}
\parbox{0.35\textwidth}{
  \includegraphics[keepaspectratio,width=0.32\textwidth]
  {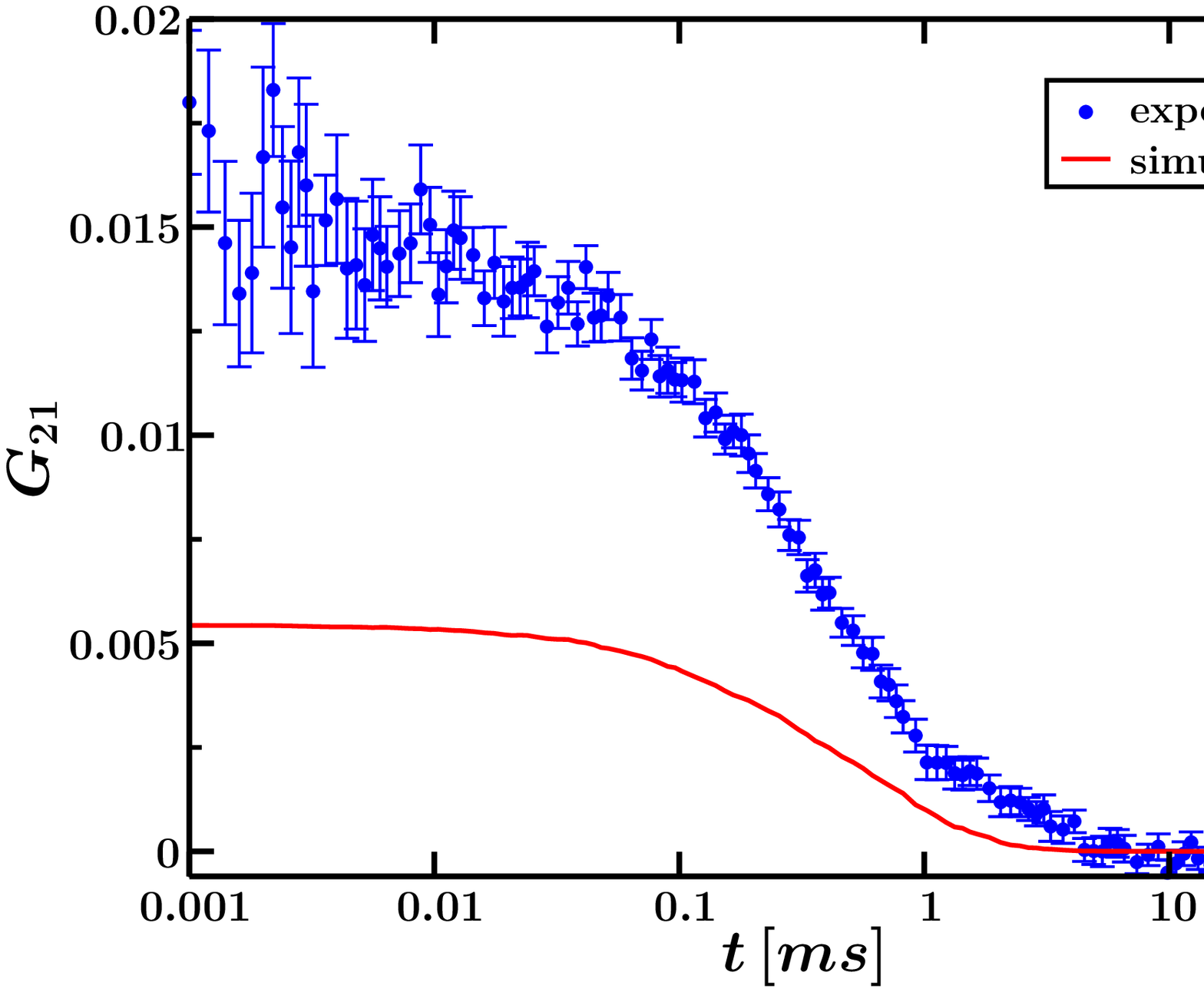} }
\\
\parbox{0.05\textwidth}{(e)}
\parbox{0.35\textwidth}{
  \includegraphics[keepaspectratio,width=0.32\textwidth]
  {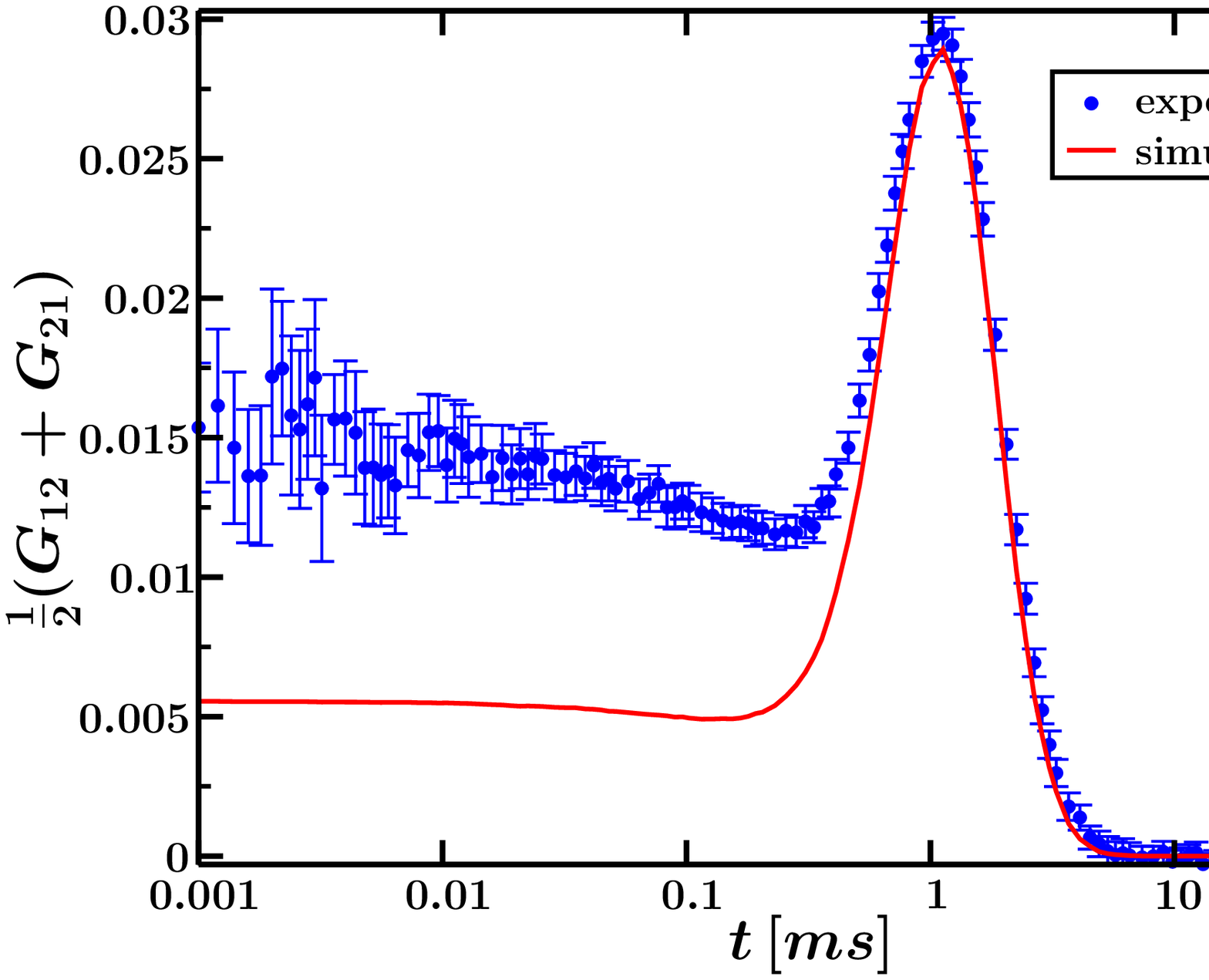} }
&
\parbox{0.05\textwidth}{(f)}
\parbox{0.35\textwidth}{
  \includegraphics[keepaspectratio,width=0.32\textwidth]
  {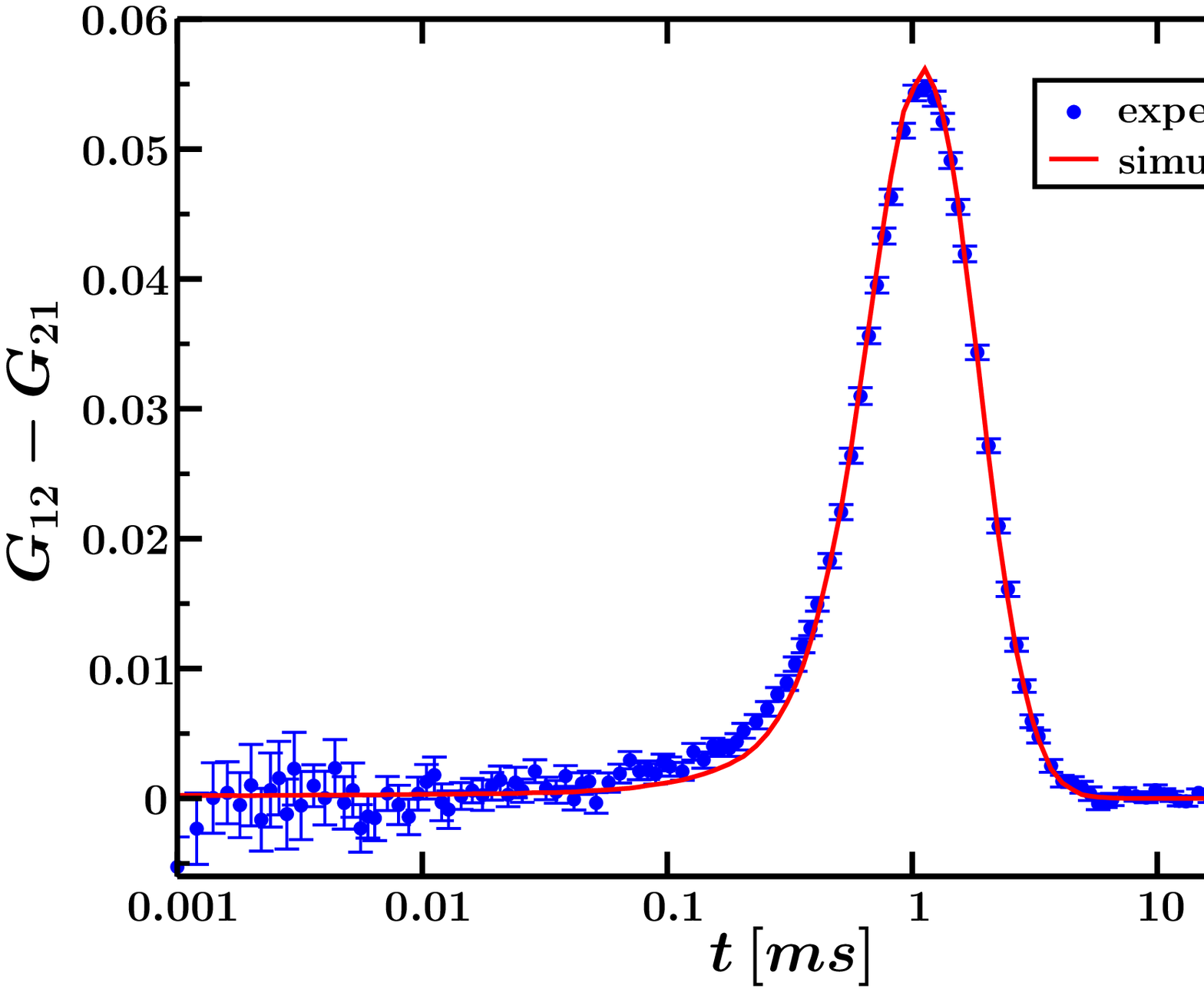} }
\\
\parbox{0.05\textwidth}{(g)}
\parbox{0.35\textwidth}{
  \includegraphics[keepaspectratio,width=0.32\textwidth]
  {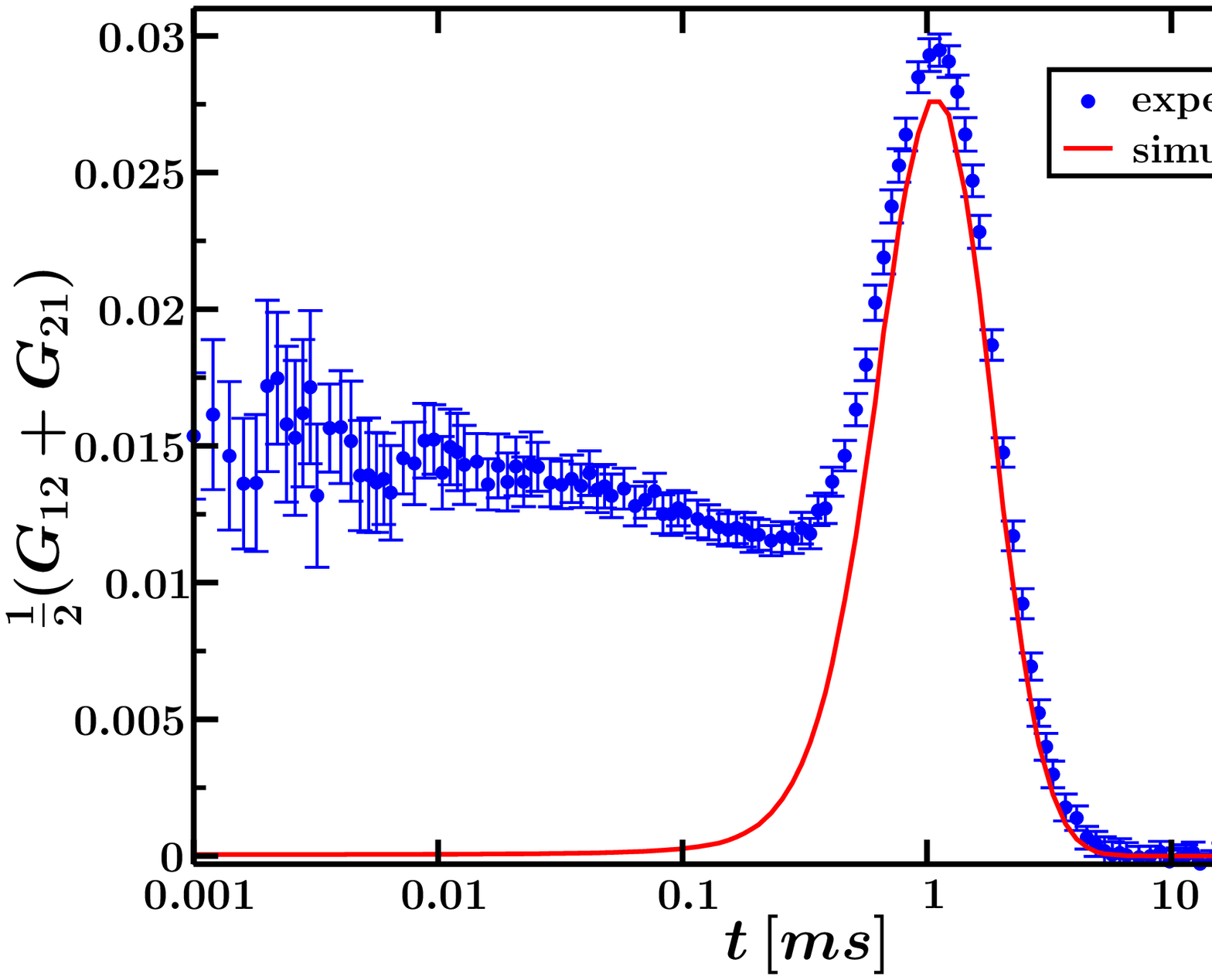} }
&
\parbox{0.05\textwidth}{(h)}
\parbox{0.35\textwidth}{
  \includegraphics[keepaspectratio,width=0.32\textwidth]
  {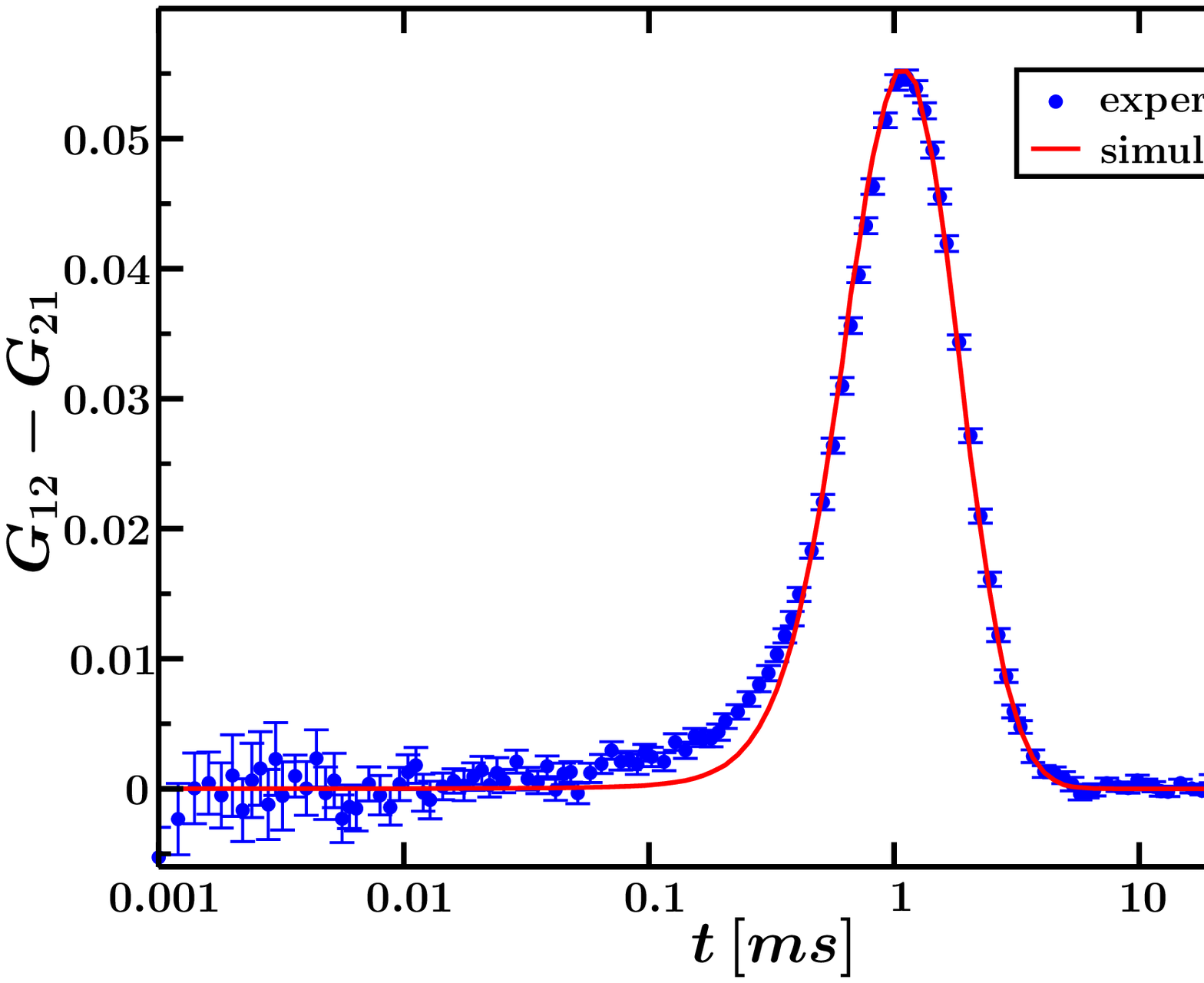} }
\end{tabular}

\caption{(Color online) Correlation functions $G_{ij}$ as defined in
  the text, and linear combinations thereof, comparing the
  experimental data (with error bars) with the numerical fit functions
  (without) for an optimized parameter set. The statistical error of
  the numerical data is smaller than the line width. Parts (a) -- (f)
  have been obtained by modeling the observation volumes by
  Eq.~\ref{eq:w1_pcpsf}, while for parts (g) and (h) we have assumed a
  Gaussian form (Eq.~\ref{eq:w1_gauss}). }
\label{fig:all_the_correlation_functions}
\end{figure*}

Figure \ref{fig:all_the_correlation_functions} summarizes our
experimental results for the $G_{ij}$ and / or linear combinations
thereof. Concerning the autocorrelation curves $G_{11}$ and $G_{22}$,
we find that they are practically identical, which means that for the
modeling it is safe to assume that both pinholes have the same
properties. This is clearly shown in part (a), where one sees that
$G_{11} - G_{22}$ differs only marginally from zero (while in our
model we have anyway strictly $G_{11} = G_{22}$). Therefore, we just
used the arithmetic mean $(G_{11} + G_{22})/2$ (part (b)) as
autocorrelation input for our fits, while we discarded the $G_{11} -
G_{22}$ data.  Concerning the cross-correlations, one sees that the
forward function $G_{12}$ (see part (c)) exhibits a pronounced peak,
which is indicative of the typical time that a particle needs to
travel from observation volume 1 to observation volume 2. Another
striking feature of $G_{12}$ is the large plateau for small times. At
such short times, the particles have essentially not moved at
all. Hence the plateau indicates that a particle is able to send
photons to both detectors from essentially the same position, or, in
other words, that the effective observation volumes must overlap quite
substantially. This overlap effect then of course also shows up in the
backward correlation function $G_{21}$ (see part (d)) at short times,
with precisely the same plateau value. Therefore, such overlap effects
essentially cancel out when considering the difference $G_{12} -
G_{21}$ instead (see part (f)), while of course they are strongly
present in the mean $(G_{12} + G_{21})/2$ (see part (e)).

Obviously, the source of the overlap must be an effect of the optical
imaging system, which is of course somewhat complicated, due to the
many components that are involved. However, beyond this general
statement we have unfortunately so far been unable to trace down its
precise physical origin, and therefore also been unable to construct a
fully consistent model for the observation volumes. The simple models
that we have considered in our present work are not fully adequate,
meaning that they systematically underestimate the amount of overlap,
unless one assumes highly unphysical parameters, which would cause
other aspects of the modeling to fail completely. It should be noted
that similar overlap effects are also present in standard double-beam
FCCS \cite{Rigler01}; however, the underlying physics for that setup
is slightly different, and the modeling used there cannot be simply
transferred to our system.

Fortunately, however, our best model for the observation volumes is at
least physical enough such that it can describe not only the
autocorrelation functions (see part (b)) but also the
overlap-corrected difference $G_{12} - G_{21}$ (part (f)) reasonably
well, while still failing to describe the mean $(G_{12} + G_{21})/2$
(part (e)). For this reason, our fitting procedure altogether takes
into account the linear combinations $(G_{11} + G_{22})/2$ and $G_{12}
- G_{21}$, while deliberately discarding the data on $(G_{12} +
G_{21})/2$ and $G_{11} - G_{22}$. This is nicely borne out in
Fig. \ref{fig:all_the_correlation_functions}, which shows not only the
experimental data, but also the result of our theoretical modeling for
optimized parameters.

The fact that the success of the modeling depends crucially on an
accurate description of the observation volumes is strongly
underpinned by parts (g) and (h) of
Fig. \ref{fig:all_the_correlation_functions}. The experimental data
for $(G_{12} + G_{21})/2$ and $G_{12} - G_{21}$ are again the same,
but the theoretical model uses a different functional form for the
observation volumes, whose performance is obviously significantly
poorer: Not only is the overlap plateau (part (g)) underestimated even
more strongly than for the better model (part (e)), but also in the
overlap-corrected function $G_{12} - G_{21}$ (part (h)) are the
deviations from the experimental data much more pronounced than for
the better model (part (f)). It should also be noted that the
autocorrelation functions are much less sensitive to these details;
the autocorrelation curve for the poorer model (data not shown) fits
the experiments as well as the better one (part (b)).

\section{Correlation Functions and Particle Dynamics}
\label{sec:corrfunc}

\subsection{Molecular Detection Efficiency}

The fluorescence particles pass consecutively through the two
observation volumes $W_1$ and $W_2$ (Fig. \ref{fig: observation volume
  plus flow}). The observation volume of each pinhole is given by the
space-dependent molecular detection efficiency (MDE) function. It
depends on the excitation intensity profile $I_z (z)$, and the
collection efficiency of the objective plus detector system. In
essence, the function $W_1(\boldsymbol{r})$ denotes the probability
density for the event that a fluorescence photon emitted from a
particle at position $\boldsymbol{r}$ will pass through pinhole $1$ 
and reach detector $1$. Similarly, $W_2(\boldsymbol{r})$ is the analogous
function for pinhole $2$. Since the intensity of the evanescent wave
decays exponentially with a penetration depth $d_p$ (of order $100
nm$), and the observation volumes are displaced with respect to one
another by a distance $s_x$ (roughly $800 nm$), we assume the
functional form
\begin{eqnarray}
W_1(\boldsymbol{r}) & = & W_{xy} \left(x, y \right)
                          d_p^{-1} \exp \left( - \frac{z}{d_p} \right) , \\
W_2(\boldsymbol{r}) & = & W_{xy} \left(x - s_x, y \right)
                          d_p^{-1} \exp \left( - \frac{z}{d_p} \right) ,
\end{eqnarray}
where normalization of the probability densities implies
\begin{equation}
\int_{-\infty}^{\infty} dx \int_{-\infty}^{\infty} dy 
W_{xy} \left(x, y \right) = 1 .
\end{equation}
In general the function $W_{xy}$ is given by the convolution of the
pinhole image in the sample space with the point spread function (PSF)
of the objective. However, one simple and widely used approximation,
valid for pinholes equal or smaller than the Airy Unit of the system,
assumes that $W_{xy}$ is a Gaussian function \cite{Rigler01,
  HasslerSecond05, zhangzerubia}:
\begin{equation} \label{eq:w1_gauss}
  W_{xy} (x,y) = \frac{2}{\pi w_0^2} 
  \exp \left( - 2 \frac{x^2 + y^2}{w_0^2} \right) ;
\end{equation}
a typical value for the width that we obtain from fitting is $w_0
\simeq 250 nm$.

\begin{figure}[t]
  \noindent
  \begin{centering}\hspace{-0.4cm}
     \includegraphics[keepaspectratio,width=0.48\textwidth]
     {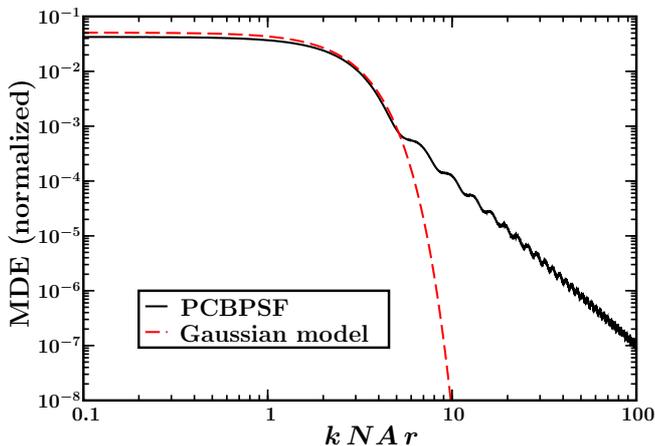}
  \end{centering}
  \caption{(Color online) Comparison of the two normalized MDEs used
    in our study, for the optimized parameters of
    Fig.~\ref{fig:all_the_correlation_functions}, using the natural
    unit system of the PCBPSF.}
  \label{fig:mde_comparison}
\end{figure}

A substantially better description of $W_{xy}$ can be obtained by
considering the explicit form of the PSF
\cite{webb,richardswolf}. However, this form is described with complex
mathematical equations and is often approximated by a squared Bessel
function \cite{webb,bornwolf,zhangzerubia}
\begin{equation}
PSF_{xy} \propto \left( \frac{2 J_1(q)}{q} \right)^2 ,
\end{equation}
where $J_1$ denotes the first Bessel function and 
\begin{equation}
q = k \, NA \, \sqrt{x^2 + y^2} =
\frac{2 \pi}{\lambda} \, NA \, \sqrt{x^2 + y^2} .
\end{equation}

Here $\lambda$ is the wavelength of the fluorescent light (in our case
$600nm$). The Bessel PSF implicitly assumes a paraxial approximation
(i.~e. small $NA$). While this assumption is probably not the best
for confocal microscopy, it is certainly more accurate than a simple
Gaussian PSF \cite{webb,zhangzerubia}.

As mentioned above, in order to describe what a pinhole sees one must
calculate the convolution of the PSF of the objective with the pinhole
image in the sample space. The geometrical image of the pinhole is
simply obtained by dividing the physical size of the pinhole (physical
radius $ = 50 \mu m$) by the total magnification of the system (in our
case $\approx 333$). This results in a radius $R_{PH}$ in the sample
space of approximately $150 nm$. Therefore the total model MDE is
given by \cite{zhangzerubia}
\begin{equation} \label{eq:w1_pcpsf}
W_{xy} (x,y) = \left( \frac{k \, NA}{2 \pi R_{PH}} \right)^2
\int_{\left\vert \boldsymbol{r_0} \right\vert \le R_{PH}} d^2 r_0 
\left( \frac{2 J_1(q)}{q} \right)^2 ,
\end{equation}
where 
\begin{equation}
q = k \, NA \, \sqrt{(x - x_0)^2 + (y - y_0)^2} .
\end{equation}
The convolution integral is difficult to evaluate analytically, but
easy to calculate numerically. To this end, we use dimensionless
length units in which the factor $k NA$ is unity. In these
dimensionless units, $R_{PH}$ takes the value $2.3$ for the parameters
given above, which is the value we have used throughout our study. We
call this function (\ref{eq:w1_pcpsf}) the ``pinhole-convoluted Bessel
point spread function'' (PCBPSF), which we calculated in dimensionless
units once and for all, and stored as a table. During the actual
data analysis, the transformation factor from dimensionless units to
real units was used as a fit parameter, in analogy to $w_0$ for the
Gaussian model. It should be noted that the PCBPSF decays for large
distances like $(x^2 + y^2)^{-3/2}$, therefore providing much more
overlap than the Gaussian model.

In the present work, we have studied both models, the ``Gaussian''
model according to Eq.~\ref{eq:w1_gauss}, as well as the PCBPSF model
according to Eq.~\ref{eq:w1_pcpsf}. The corresponding correlation
curves have already been presented in
Fig.~\ref{fig:all_the_correlation_functions}. The corresponding MDEs
are shown in Fig.~\ref{fig:mde_comparison}. One sees that the PCBPSF
model puts much more statistical weight into the tail of the
distribution than the Gaussian model. As already discussed above, we
found the Gaussian model to perform less well than the PCBPSF model,
since it underestimates the overlap even more severely than the
latter. In what follows, we will present data always for the PCBPSF
model, unless stated differently.

\subsection{Theory of Correlation Functions}

The dynamics of the tracer particles is described by the space- and
time-dependent concentration (number of particles per unit volume)
$C (\boldsymbol{r},t)$, its fluctuation
\begin{equation}
\delta C (\boldsymbol{r},t) = C (\boldsymbol{r},t) - \langle C \rangle
\end{equation}
and the concentration correlation function
\begin{equation}
   \Phi ({\boldsymbol{r}}, {\boldsymbol{r}}', t) =
   \langle \delta C({\boldsymbol{r}}, t) 
   \delta C({\boldsymbol{r}}', 0) \rangle ;
\end{equation}
note that translational invariance applies only to time, but not to
space, due to the presence of the flow and the surface. At time $t =
0$, this reduces to the static correlation function, for which we
simply assume the function pertaining to an ideal gas:
\begin{equation} \label{eq:ideal gas initial condition}
\Phi ({\boldsymbol{r}}, {\boldsymbol{r}}', 0) =
\langle C \rangle \delta( {\boldsymbol{r}} - {\boldsymbol{r}}') .
\end{equation}
Note that this assumption implies that we consider the particles as
point particles, with no interaction with the surface except
impenetrability, and no interaction between each other, due to
dilution.

As described in Ref. \cite{Rigler01}, the correlation functions are
related to $\Phi$ via
\begin{eqnarray} \label{eq: double integral}
   G_{ij}(t) & = & \frac{\int \int d^3 r d^3 r' W_i({\boldsymbol{r}'})
             W_j({\boldsymbol{r}})
             \Phi ({\boldsymbol{r}}, {\boldsymbol{r}'}, t)}
             { \langle C \rangle ^2 \left( 
               \int d^3 r W_i({\boldsymbol{r}}) \right) 
             \left( \int d^3 r W_j({\boldsymbol{r}}) \right) }
\\
\nonumber
   & = & \langle C \rangle^{-2} 
   \int \int d^3 r d^3 r' W_i({\boldsymbol{r}'})
   W_j({\boldsymbol{r}})
   \Phi ({\boldsymbol{r}}, {\boldsymbol{r}'}, t) ,
\end{eqnarray}
where in the second step we have taken into account the normalization
of the $W_i$. Therefore, the obvious strategy for analyzing the
experimental data is to (i) evaluate $\Phi$ within a model for the
particle dynamics, (ii) evaluate the integrals in Eq. \ref{eq: double
  integral} to obtain a theoretical prediction for $G_{ij}$ for a
given set of parameters, (iii) compare the prediction with the data,
and (iv) optimize the parameters. The normalizing prefactor $\langle
C \rangle^{-2}$ is not known very accurately and will hence be treated
as a fit parameter.

The tracer particles undergo a diffusion process and move in an
externally driven flow field $\boldsymbol{v}$. Hence, we describe the
concentration correlation function by a convection-diffusion equation
of the form
\begin{equation} \label{eq: CDE}
   \partial_t \Phi ({\boldsymbol{r}}, {\boldsymbol{r}}', t)
   = D \nabla_{\boldsymbol{r}}^2 
     \Phi ({\boldsymbol{r}}, {\boldsymbol{r}}', t)
   - \nabla_{\boldsymbol{r}} \cdot {\boldsymbol{v} (\boldsymbol{r}) }
   \Phi ({\boldsymbol{r}}, {\boldsymbol{r}}', t),
\end{equation}
which needs to be solved for $z \ge 0, z' \ge 0$ with the initial
condition Eq. \ref{eq:ideal gas initial condition} and the no-flux
boundary condition at the surface,
\begin{equation}
\left. \partial_z \Phi ({\boldsymbol{r}}, {\boldsymbol{r}}', t)
\right\vert_{z=0} = 0 ,
\end{equation} 
which imposes that there is no diffusive current entering the solid.
For reasons of simplicity, the hydrodynamic interactions with the
surface are neglected, and hence the diffusive term is described only
by an isotropic diffusion constant $D$. 

Since in the experiment the exponential decay length of the spatial
detection volume normal to the surface is in the range of $100 - 200
nm$, while the channel size is three orders of magnitude larger, it is
justified to assume the flow field to be approximately linear. For our
geometry, this implies
\begin{equation} \label{eq: our linear flow field}
  \boldsymbol{v} (\boldsymbol{r}) = 
  \dot \gamma \tensor{\boldsymbol{\varepsilon}} \cdot
  (\boldsymbol{r} + l_s \hat{\boldsymbol{e}}_z),
\end{equation}
where $l_s$ is the slip length, $\dot \gamma = \partial v_x / \partial
z$ is the constant shear rate, $\hat{\boldsymbol{e}}_z$ denotes the
unit vector in $z$-direction and $\tensor{\boldsymbol{\varepsilon}} =
\hat{\boldsymbol{e}}_x \otimes \hat{\boldsymbol{e}}_z$ is the
dimensionless rate-of-strain tensor.

At this point, it is useful to re-define the coordinate system in such
a way that the finite hard-sphere radius $R$ of the tracer particles
(roughly $7 nm$) is taken into account. We therefore identify $z = 0$
no longer with the interface, but rather with the $z$ coordinate
of the particle center at contact with the interface. In this new
coordinate system, the flow field is given by
\begin{equation} \label{eq: new linear flow field}
  \boldsymbol{v} (\boldsymbol{r}) = 
  \dot \gamma \tensor{\boldsymbol{\varepsilon}} \cdot
  (\boldsymbol{r} + (l_s + R) \hat{\boldsymbol{e}}_z) ,
\end{equation}
i.~e. we simply have to add the particle radius to the slip length.
The functional form of the observation volumes $W_1$ and $W_2$ remains
unchanged, since the $z$ dependence is just an exponential decay, such
that a shift in $z$ direction just results in a constant prefactor
that can be absorbed in the overall normalization. Our method therefore
does not yield a value for $l_s$, but rather only for the combination
$l_s + R$.

As mentioned previously, for some special cases the
convection-diffusion equation can be solved analytically, for example
in the case of uniform or linear flow in bulk, i.~e. far away from
surfaces \cite{Lum03, Bri99, Elr62, Foi80}, or for pure diffusion
close to the wall, but without any flow field \cite{Rie08,
  Hassler05}. For our case, however, it is not easy, or even
impossible, to find such a solution. Therefore the aim of the next
sections will be to construct a stochastic numerical
method. Concerning the problems that were mentioned after Eq. \ref{eq:
  double integral}, (i) and (ii) can be solved by Brownian Dynamics,
while problems (iii) and (iv) are tackled by a Monte Carlo algorithm
in parameter space.

\section{Sampling algorithm}
\label{Sec: Sampling Algorithm}

Brownian motion of particles under the influence of external driving
is described by a Fokker-Planck equation \cite{Ris96, Hon94,
  Oet96, Maz02, Lad03}, which has exactly the same form as the
convection-diffusion equation, Eq.  \ref{eq: CDE}, the only difference
being that $\Phi$ is replaced by the so-called ``propagator''
$P (\boldsymbol{r}, t \vert {\boldsymbol{r}}', 0)$, which is the
conditional probability density for the particle motion $\boldsymbol{r}'
\to \boldsymbol{r}$ within the time $t$. $P$ and $\Phi$ describe the
same physics and are actually identical except for a trivial normalization
factor, $\Phi = \langle C \rangle P$. We can therefore rewrite
Eq.~\ref{eq: double integral} as
\begin{eqnarray} \label{eq:double_integral_simplified}
 & & \langle C \rangle G_{ij} (t) \\
\nonumber
 & = &
   \int \int d^3 r d^3 r' W_i({\boldsymbol{r}'})
   W_j({\boldsymbol{r}})
   P (\boldsymbol{r}, t \vert {\boldsymbol{r}}', 0) .
\end{eqnarray}
As is well-known, the Fokker-Planck equation is equivalent to
describing the particle dynamics in terms of a Langevin equation
\begin{equation} \label{eq: Langevin}
  \dot{\boldsymbol{r}} (t) =
  {\boldsymbol{v}}({\boldsymbol{r}}(t)) + {\boldsymbol{\eta}}(t) .
\end{equation}
Here $\dot{\boldsymbol{r}} (t)$ is the tracer velocity,
${\boldsymbol{v}}$ is the deterministic (external) velocity imposed by
the flow, while ${\boldsymbol{\eta}}$ is a stochastic Gaussian white
noise term which describes the diffusion:
\begin{subequations}\label{eq: noise moments}
   \begin{eqnarray}
     \langle \eta_\alpha (t) \rangle & = & 0, \\
     \langle \eta_\alpha (t') \eta_\beta (t) \rangle
     & = & 2 D \delta_{\alpha \beta} \delta(t'-t) .
   \end{eqnarray}
\end{subequations}
Here, $\alpha,\beta = x,y,z$ are Cartesian indices and $\delta_{\alpha
  \beta}$ is the Kronecker delta. We solve this Langevin equation
numerically by means of a simple Euler algorithm \cite{Oet96} with a
finite time step $\Delta t$:
\begin{equation} \label{eq: Euler SI}
  \boldsymbol{r} (t + \Delta t) = \boldsymbol{r} (t)
   + \Delta t {\boldsymbol{v}}({\boldsymbol{r}}(t))
   + \sqrt{2 D \Delta t}  {\boldsymbol{\chi}} ,
\end{equation}
where $\boldsymbol{\chi} = (\chi_x, \chi_y, \chi_z)$ is a vector of
mutually independent random numbers with mean $0$ and variance $1$.
The boundary condition at the wall is taken into account by a simple
reflection at $z = 0$, i.~e. a particle that, after a certain time
step, has entered the negative half-space $z < 0$ is subjected to
$z \to - z$ before the next propagation step is executed.

Now, let us consider a computer experiment where, at time $t = 0$, we
place a particle randomly in space, with probability density $\rho_0(
\boldsymbol{r'} )$, and then propagate it stochastically according to
Eq. \ref{eq: Euler SI}. The probability density for it reaching the
position $\boldsymbol{r}$ after the time $t$ is then given by
\begin{equation}
Q( \boldsymbol{r}, t ) =
\int d^3 r' P (\boldsymbol{r}, t \vert {\boldsymbol{r}}', 0)
\rho_0( \boldsymbol{r'} ) .
\end{equation}

If we now consider an observable $A$, which is some function of the
particle's coordinate, $A = A( \boldsymbol{r} )$, and study the time
evolution of its average, then this is obviously given by
\begin{eqnarray}
\langle A \rangle (t) 
& = &
\langle A ( \boldsymbol{r} (t) ) \rangle \\
\nonumber
& = &
\int d^3 r  A ( \boldsymbol{r} ) Q( \boldsymbol{r}, t ) \\
\nonumber
& = &
\int \int d^3 r d^3 r' A ( \boldsymbol{r} )
 P (\boldsymbol{r}, t \vert {\boldsymbol{r}}', 0)
\rho_0( \boldsymbol{r'} ) .
\end{eqnarray}

\begin{figure}[t]
   \noindent
   \begin{centering}\hspace{-0.4cm}
       \includegraphics[clip,width=0.49\textwidth]
       {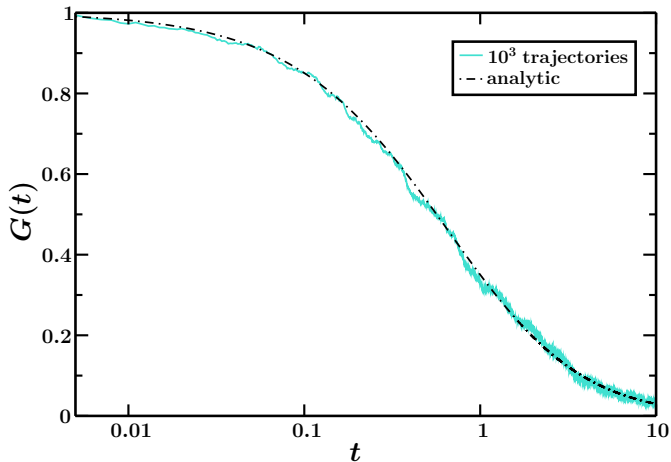}
   \end{centering}
   \caption{(Color online) Analytical solution and simulated data for
     an average over $10^3$ trajectories.}
   \label{fig: analytic vs simulation}
\end{figure}

Therefore, if we set $\rho_0 = W_i$ and $A = W_j$, then $\left< A
\right>$ is identical to the rescaled correlation function $\langle C
\rangle G_{ij}$. In other words, we place the particle initially with
probability density $W_i$, then generate a stochastic trajectory via
Eq. \ref{eq: Euler SI}, and evaluate $W_j$ for all times along that
trajectory. This yields a function $W_j (t)$ for that particular
trajectory. This computer experiment is repeated often, and averaging
$W_j(t)$ over all trajectories yields directly a stochastic estimate
for the (unnormalized) correlation function $G_{ij}$. Of course, these
estimates will have statistical error bars, just as the experimental
ones; however, we sample several hundred thousand trajectories, such
that the numerical errors are substantially smaller than the
experimental ones. In principle, the numerical data are also subject
to a systematic discretization error as a result of the finite time
step; however, by choosing a small value for $\Delta t$ we have made
sure that this is still small compared to the statistical
uncertainty. Note also that our approach implements an optimal
importance sampling \cite{Lan00} with respect to the $t = 0$ factor
$W_i$, but not with respect to $W_j$. In practical terms, our
straightforward sampling scheme turned out to be absolutely adequate.

The simulations were run using a ``natural'' unit system where length
units are defined by setting $d_p$ to unity, while the time units are
given by setting the diffusion constant $D$ to unity. The time step
was fixed in physical units to a value of at most $2 \mu s$ (it was
dynamically adjusted in order to match the non-equidistant
experimental observation times), which, for all parameters, is much
smaller than unity in dimensionless units. Obviously, this is small
enough to represent the stochastic part of the Langevin update scheme
with sufficient accuracy. For typical parameters ($D = 35 \mu m^2 /
s$, $d_p = 0.1 \mu m$, $\dot \gamma = 4 \times 10^{3} s^{-1}$), the
dimensionless unit time corresponds to $\simeq 0.3 ms$, such that the
resulting value for the dimensionless shear rate ($\simeq 1.2$) is of
order unity as well. Since $d_p$ (or unity, in dimensionless units)
defines the $z$ range in which the statistically relevant part of the
simulation takes place, we find that typical flow velocities in
dimensionless units are also of order unity. This shows that the time
step is also small enough for the deterministic part of the Langevin
equation. We also see that the experiment is neither dominated by
diffusion nor by convection, and therefore the analysis needs to take
into account both.

\begin{figure}[t]
   \noindent
   \begin{centering}\hspace{-0.4cm}
      \includegraphics[clip,width=0.49\textwidth]
      {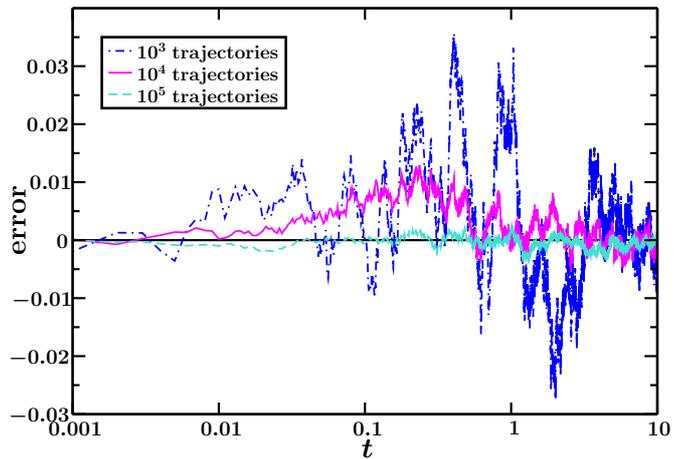}
   \end{centering}
   \caption{(Color online) Deviation from the analytic curve for
     $10^3$,$10^4$ and $10^5$ trajectories.}
    \label{fig: analytic vs simulation error}
\end{figure}

As a simple test case, we used our algorithm to calculate the
autocorrelation function for vanishing flow and the Gaussian model for
the observation volume, where an analytical solution is known
\cite{Hassler05, Rie08}. In our dimensionless units, it is, up to a
constant prefactor, given by
\begin{eqnarray} \label{eq: Analytic Solution}
   & & G^{(\mathrm{a})}(t )  \\
   \nonumber
   & = &
   \left( 1 + \frac{4 t}{w_0^2} \right)^{-1} \left(
   \left(1 - 2 t \right) \exp\left( t \right)
   \mathrm{erfc} \left[ \sqrt{t} \right]
   + \sqrt{\frac{4}{\pi} t} \right) .
\end{eqnarray}
Figure \ref{fig: analytic vs simulation} shows the analytic
autocorrelation function with $w_0 = 2$ and its simulated counterpart,
averaged over $10^3$ independent trajectories, where a small time step
of $\Delta t = 10^{-3}$ (in dimensionless units) was used. In
Fig. \ref{fig: analytic vs simulation error} the deviation of the
simulated data ($G^{(\mathrm{s})}$) from the analytic expression is
shown,
\begin{equation}
  \mathrm{error} (t) = G^{(\mathrm{s})}(t) - G^{(\mathrm{a})}(t) .
\end{equation} 
Clearly, the numerical solution converges to the analytical result
when the number of trajectories is increased, as it should be.

\section{Subtraction Scheme}
\label{sec:subtraction}

At this point, it is worthwhile to reconsider the subtraction
procedure introduced in Sec. \ref{sec:exp_corrfunc}. To this end, we
assume that the true functions $W_i$ differ somewhat from the model
functions, which we will denote by $W_i^{(m)}$. This is most
easily parameterized by the ansatz
\begin{equation} \label{eq:wi_deviation}
W_i = 
\left( 1 - \varepsilon \right) W_i^{(m)} + \varepsilon \tilde W_i ,
\end{equation}
where $W_i$, $W_i^{(m)}$ and $\tilde W_i$ are all normalized to unity,
while $\varepsilon$ is a (hopefully) small parameter. For the purposes
of the present analysis, we also assume that the Brownian Dynamics
model is a faithful and correct description of the true dynamics,
i.~e. that the difference between $W_i$ and $W_i^{(m)}$ is the only
reason for a systematic deviation between simulation and experiment.

Inserting Eq. \ref{eq:wi_deviation} into
Eq. \ref{eq:double_integral_simplified}, we thus find
\begin{eqnarray}
 & & \langle C \rangle G_{ij} (t)
\\
\nonumber
 & = &
   \left( 1 - \varepsilon \right)^2 
   \int \int d^3 r d^3 r' W_i^{(m)} ({\boldsymbol{r}'})
   W_j^{(m)} ({\boldsymbol{r}})
   P (\boldsymbol{r}, t \vert {\boldsymbol{r}}', 0) 
\\
\nonumber
 & + &
   \varepsilon \left( 1 - \varepsilon \right)
   \int \int d^3 r d^3 r' W_i^{(m)} ({\boldsymbol{r}'})
   \tilde W_j({\boldsymbol{r}})
   P (\boldsymbol{r}, t \vert {\boldsymbol{r}}', 0)
\\
\nonumber
 & + &
   \varepsilon \left( 1 - \varepsilon \right)
   \int \int d^3 r d^3 r' \tilde W_i({\boldsymbol{r}'})
   W_j^{(m)} ({\boldsymbol{r}})
   P (\boldsymbol{r}, t \vert {\boldsymbol{r}}', 0)
\\
\nonumber
 & + &
   \varepsilon^2
   \int \int d^3 r d^3 r' \tilde W_i({\boldsymbol{r}'})
   \tilde W_j({\boldsymbol{r}})
   P (\boldsymbol{r}, t \vert {\boldsymbol{r}}', 0) .
\end{eqnarray}
Since we treat $\langle C \rangle$ as an adjustable parameter, it
makes sense to view the first term (including the prefactor $\left( 1
  - \varepsilon \right)^2$) as the theoretical model for the
correlation function, $\left< C \right> G_{ij}^{(m)} (t)$. For the
deviation between experiment and theory we then obtain, neglecting all
terms of order $\varepsilon^2$,
\begin{eqnarray} \label{eq:kij_orig}
K_{ij} & := &
\varepsilon^{-1} \langle C \rangle \left( G_{ij} - G_{ij}^{(m)} \right)
\\
\nonumber
& = &
   \int \int d^3 r d^3 r' 
   W_i^{(m)} ({\boldsymbol{r}'}) \tilde W_j ({\boldsymbol{r}})
   P (\boldsymbol{r}, t \vert {\boldsymbol{r}}', 0)
\\
\nonumber
& + &
  \int \int d^3 r d^3 r' 
   \tilde W_i({\boldsymbol{r}'}) W_j^{(m)} ({\boldsymbol{r}})
   P (\boldsymbol{r}, t \vert {\boldsymbol{r}}', 0) ,
\end{eqnarray}
and for its antisymmetric part
\begin{eqnarray}
&& K_{ij} - K_{ji}
\\
\nonumber
& = &
  \int \int d^3 r d^3 r' [
  W_i^{(m)} ({\boldsymbol{r}'}) \tilde W_j ({\boldsymbol{r}}) 
\\
\nonumber
&&
- W_i^{(m)} ({\boldsymbol{r}}) \tilde W_j ({\boldsymbol{r}'})
  ]
  P (\boldsymbol{r}, t \vert {\boldsymbol{r}}', 0)
\\
\nonumber
& - &
  \int \int d^3 r d^3 r' [
  W_j^{(m)} ({\boldsymbol{r}'}) \tilde W_i ({\boldsymbol{r}})
\\
\nonumber
&&
- W_j^{(m)} ({\boldsymbol{r}}) \tilde W_i ({\boldsymbol{r}'})
  ]
  P (\boldsymbol{r}, t \vert {\boldsymbol{r}}', 0) .
\end{eqnarray}
The terms in square brackets are antisymmetric under the exchange
$\boldsymbol{r} \leftrightarrow \boldsymbol{r}'$, and hence $P$
can be replaced by its antisymmetric part
\begin{equation}
P_a (\boldsymbol{r}, {\boldsymbol{r}}', t) =
P (\boldsymbol{r}, t \vert {\boldsymbol{r}}', 0) -
P (\boldsymbol{r}', t \vert {\boldsymbol{r}}, 0) .
\end{equation}
Exchanging the arguments in the second terms within the square
brackets then yields
\begin{eqnarray} \label{eq:subtract_anal_final_result}
&& \frac{1}{2} \left( K_{ij} - K_{ji} \right)
\\
\nonumber
& = &
  \int \int d^3 r d^3 r'
  W_i^{(m)} ({\boldsymbol{r}'}) \tilde W_j ({\boldsymbol{r}}) 
  P_a (\boldsymbol{r}, {\boldsymbol{r}}', t)
\\
\nonumber
& - &
  \int \int d^3 r d^3 r'
  W_j^{(m)} ({\boldsymbol{r}'}) \tilde W_i ({\boldsymbol{r}})
  P_a (\boldsymbol{r}, {\boldsymbol{r}}', t) .
\end{eqnarray}
This is clearly a nonzero contribution. In other words, the
subtraction scheme (i.~e. studying $G_{12} - G_{21}$ instead of
$G_{12}$) does \emph{not} provide a consistent cancellation procedure
such that the first-order deviation would vanish. However, in
practical terms the deviation is much smaller than for the original
data ($G_{12}$ and $G_{21}$), for which Eq.~\ref{eq:kij_orig} applies.
To some extent, this is so because the error is the difference of two
terms, but mostly it is due to the fact that not the full propagator
$P$ contributes, but rather only its antisymmetric part $P_a$. For
short times the dynamics is dominated by diffusion, i.~e. $P$ is
essentially symmetric, or $P_a \approx 0$. At late times, we again
expect $P_a$ to become quite small (exponentially damped, see
Eq.~\ref{eq:antisymmetric_propagator}), although we have no rigorous
proof for this.  Therefore one should expect that the strongest
deviation occurs at intermediate times where $P_a$ is maximum. This
time scale is not given by the optical geometry but rather by the
dynamics; dimensional analysis then tells us that this time must be of
order $D / v^2$.  For typical parameters of our experiment ($D = 35
\mu m^2 / s$, $v = 4 \times 10^2 \mu m / s$) we obtain a value of
roughly $0.2 ms$, which fits quite well to the observations one can
make in Fig.~\ref{fig:all_the_correlation_functions}, part (h).  At
such times, we expect that the main contribution to $K_{12} - K_{21}$
comes form the first term of Eq.~\ref{eq:subtract_anal_final_result}
(downstream vs. upstream correlation) and that $P_a$ is positive for
most of the relevant arguments. Therefore, one should expect that the
experimental data should lie systematically above the theoretical
predictions, which is indeed the case. Our expectations concerning the
behavior of $P_a$ come from studying the simple case of
one-dimensional diffusion with constant drift without boundary
conditions; here one has
\begin{equation}
P(x,t \vert x',0) = \frac{1}{\sqrt{4 \pi D t}}
\exp \left( - \frac{(x-x'-vt)^2}{4 D t} \right)
\end{equation}
and
\begin{eqnarray} \label{eq:antisymmetric_propagator}
P_a(x,x',t) & = & \frac{2}{\sqrt{4 \pi D t}}
\exp \left( - \frac{(x-x')^2}{4 D t} \right)
\\
\nonumber
&&
\exp \left( - \frac{v^2 t}{4 D} \right)
\sinh \left( \frac{ (x-x') v}{2 D} \right) .
\end{eqnarray}

\section{Statistical Data Analysis}
\label{sec: stat_dat}

\subsection{Monte Carlo Algorithm}

For the model that we consider in the present paper, the space of fit
parameters is (in principle) seven-dimensional. We have three lengths
that define the geometry of the optical setup, $d_p$, $s_x$, and $w_0$
(Gaussian model) or $(k \, NA)^{-1}$ (diffraction model). Three
further parameters define the properties of the flow and the diffusive
dynamics of the tracers; these are the diffusion constant $D$, the
shear rate $\dot \gamma$, and the slip length plus particle radius
$l_s + R$. Finally, there is the concentration of tracer particles
$\langle C \rangle$, which serves as a global normalization
constant. The functions to be fitted are $(G_{11} + G_{22}) / 2$ and
$G_{12} - G_{21}$. However, we have seen in Sec.~\ref{sec:subtraction}
that the non-idealities in modeling the observation volumes do have an
effect on the normalizations, and therefore we allowed one separate
normalization constant $\langle C \rangle$ for each of the curves
($\langle C \rangle_A$ for the autocorrelation and $\langle C
\rangle_C$ for the cross-correlation), in order to partly compensate
for these non-idealities. Therefore, our parameter space is finally
eight-dimensional. The strategy that we develop in the present section
aims at adjusting all parameters simultaneously in order to obtain
optimum fits. For the further development, it will be useful to
combine all the parameters into one vector
$\boldsymbol{\Pi}$. Furthermore, for each parameter we can, from
various physical considerations, define an interval within which it is
allowed to vary (because values outside that interval would be highly
unreasonable or outright unphysical). This means that we restrict the
consideration to a finite eight-dimensional box
$\Omega_{\boldsymbol{\Pi}}$ in parameter space.

A central ingredient of our approach is the fact that both the
experimental data and the simulation results have been obtained with
good statistical accuracy ($\simeq 2.5 \times 10^5$ trajectories for
the simulations, $40$ independent measurements for the
experiments). This does not only allow us to obtain rather small
statistical error bars, but also (even more importantly) to rely on
the asymptotics of the Central Limit Theorem, i.~e. to assume Gaussian
statistics throughout. For both correlation curves and each of the
considered times, we have both an experimental data point $E_i$ and a
simulated data point $S_i$, where the index $i$ simply enumerates the
data points. Both $E_i$ and $S_i$ can be considered as Gaussian random
variables with variances $\sigma_{E,i}^2$ and $\sigma_{S,i}^2$,
respectively. Then
\begin{equation}
\tilde \Delta_i 
= \frac{S_i - E_i}{ \sqrt{ \sigma_{S,i}^2 + \sigma_{E,i}^2 } }
\end{equation}
is again a Gaussian random variable, whose variance is simply unity,
\begin{equation}
\left< \tilde \Delta_i^2 \right> - \left< \tilde \Delta_i \right>^2 = 1 .
\end{equation}
Therefore, $\tilde \Delta_i$ is, in principle, a perfect variable to
measure the deviation between simulation and experiment.
Unfortunately, however, the parameters $\sigma_{S,i}$ and
$\sigma_{E,i}$ are not known. What is rather known are their
\emph{estimators} $s_{S,i}$ and $s_{E,i}$, as they are obtained from
standard analysis to calculate error bars. Therefore, we rather
consider
\begin{equation}
\Delta_i 
= \frac{S_i - E_i}{ \sqrt{ s_{S,i}^2 + s_{E,i}^2 } } .
\end{equation}
The statistical properties of this variable, however, are in the
general case unknown \cite{Wel38}. It is only in the case of rather
good statistics (as we have realized it) that we can ignore the
difference between $\sigma$ and $s$, and simply assume that $\Delta_i$
is indeed a Gaussian variable with unit variance. It is at this point
where the statistical quality of the data clearly becomes important.

If $M$ is the total number of data points, then
\begin{equation}
{\cal H} = \frac{1}{2} \sum_{i = 1}^M \Delta_i^2
\end{equation}
is obviously a quantity that measures rather well the deviation
between experiment and simulation. In principle, the task is to pick
the parameter vector $\boldsymbol{\Pi}$ in such a way that ${\cal H}$
is minimized. We have deliberately chosen the symbol ${\cal H}$ in
order to point out the analogy to the problem of finding the ground
state of a statistical-mechanical Hamiltonian. In case of a perfect
fit, we have $\left< S_i \right> = \left< E_i \right>$ or $\left<
  \Delta_i \right> = 0$, implying $\left< {\cal H} \right> = M/2$. In
the standard nomenclature of fitting problems, $2 {\cal H}$ is called
``chi squared''. We also introduce $\xi = 2 {\cal H} / M$, which we
will call the ``goodness of simulation'' (standard nomenclature: ``chi
squared per degree of freedom'').

For optimizing $\boldsymbol{\Pi}$, we obviously need to consider
${\cal H}$ as a function of $\boldsymbol{\Pi}$. In this context, it
turns out that it is important to be able to consider it as a
function of \emph{only} $\boldsymbol{\Pi}$, and to make sure that this
dependence is smooth. For this reason, we use the \emph{same} number
of trajectories when going from one parameter set to another one, and
use \emph{exactly the same} set of random numbers to generate the
trajectories. In other words, the trajectories differ \emph{only} due
to the fact that the parameters were changed. Therefore, both $S_i$
and $s_{S,i}$ are smooth functions of the parameters, and ${\cal H}$
is as well.

In order to find the optimum parameter set, one could, in principle,
construct a regular grid in $\Omega_{\boldsymbol{\Pi}}$ and then
evaluate ${\cal H}$ for every grid point. However, for
high-dimensional spaces (and eight should in this context be viewed as
already a fairly large number, in particular when taking into account
that it is bound to increase further as soon as more refined models
are studied), it is usually more efficient to scan the space by an
importance-sampling Monte Carlo procedure based upon a Markov chain
\cite{Lan00}. Applying the standard Metropolis scheme \cite{Lan00}, we
thus arrive at the following algorithm:
\begin{enumerate}
\item Choose some start vector $\boldsymbol{\Pi}$. This should
      be a reasonable set of parameters, perhaps pre-optimized
      by simple visual fitting.
\item From the previous set of parameters, generate a trial
      set via $\boldsymbol{\Pi}' = \boldsymbol{\Pi} + 
      \Delta \boldsymbol{\Pi}$, where $\Delta \boldsymbol{\Pi}$
      is a random vector chosen from a uniform distribution
      from a small sub-box aligned with $\Omega_{\boldsymbol{\Pi}}$.
\item If the new vector is not within $\Omega_{\boldsymbol{\Pi}}$,
      reject the trial set and go to step 2.
\item Otherwise, calculate both $P_{eq} (\boldsymbol{\Pi}')$
      and $P_{eq} (\boldsymbol{\Pi})$, as well as the Metropolis
      function
      \begin{equation}
       m = \min \left(1, 
       P_{eq} (\boldsymbol{\Pi}') / P_{eq} (\boldsymbol{\Pi}) \right) ,
      \end{equation}
      where $P_{eq}$ is the ``equilibrium'' probability density
      of $\boldsymbol{\Pi}$, i.~e. the desired probability density
      towards which the Markov chain converges (more about this below).
\item Accept the trial move with probability $m$ (reject it with
      probability $1 - m$), count either the
      accepted or the old set as a new set in the Markov chain,
      and go to step 2.
\item After relaxation into equilibrium, sample desired properties
      of the distribution of $\boldsymbol{\Pi}$, like mean values,
      variances, covariances, etc., by simple arithmetic means
      over the parameter sets that have been generated by the Markov
      chain. This allows the estimation of not only the physical parameters,
      but at the same time also of their statistical error bars.
\end{enumerate}
The scheme is defined as soon as $P_{eq}$ is specified. Now, from the
considerations above, we know that \emph{in case of a perfect fit} the
variables $\Delta_i$ are independent Gaussians with zero mean and unit
variance. This implies (ignoring constant prefactors which anyway
cancel out in the Metropolis function)
\begin{eqnarray}
\nonumber
P_{eq} & \propto & \prod_i \exp \left( - \frac{1}{2} \Delta_i^2 \right) \\
\nonumber
      & = & \exp \left( - \frac{1}{2} \sum_i \Delta_i^2 \right) \\
      & = & \exp \left( - {\cal H} \right) , \
\end{eqnarray}
which makes the interpretation in terms of statistical mechanics
obvious. Clearly, this form for $P_{eq}$ is the only reasonable choice
for implementing the Monte Carlo algorithm. After relaxation into
equilibrium, one should observe a $\xi$ value of roughly unity, while
larger numbers indicate a non-perfect fit (even after exhaustive
Monte Carlo search), and thus deficiencies in the theoretical model.
One should also be aware that the equilibrium fluctuations of $\xi$ are
expected to be quite small, since $\xi$ is the arithmetic mean of a fairly
large number ($M$, the number of experimental data points) of independent
random variables.

In practice, we adjusted $\Delta \boldsymbol{\Pi}$ in order to obtain
a fairly large acceptance rate of roughly $0.6 \ldots 0.8$. The Monte
Carlo algorithm was then run for more than $3 \times 10^5$ steps, each
step involving the generation of roughly $2.5 \times 10^5$
trajectories. The simulation was run on $512$ nodes ($2048$ processes)
of the IBM Blue Gene-P at Rechenzentrum Garching, where each process
generated $123$ trajectories. On this machine, one Monte Carlo run
took roughly one day to complete. It turned out that discarding the
first $5 \times 10^4$ configurations was sufficient to obtain data in
equilibrium conditions, where the mean values of the parameters and
their standard deviations were calculated. It should be noted that the
equilibrium fluctuations of the parameters tell us the typical range
in which they can still be viewed as compatible with the
experiments. Therefore these fluctuations are the appropriate measure
to quantify the experimental error bars, while calculating a standard
error of mean (or a similar quantity) would \emph{not} be appropriate
and severely underestimate the errors. Finally, it should be noted
that the approach allows in principle to analyze the mutual dependence
of the parameters as well, by sampling the corresponding covariances;
this was however not done in the present study.

\subsection{Scale Invariance}

As noted before, the correlation functions depend on the average
concentration $\langle C \rangle$, the diffusion constant $D$, the
shear rate $\dot \gamma$, and various lengths, which we denote by
$\{l_i\}$. Simple dimensional analysis shows that for any
scale factor $a$ the scaling relation
\begin{eqnarray}
  && G_{ij}( t, \langle C \rangle, D, \dot \gamma, \{l_i\} ) \\
  \nonumber
  & = & G_{ij}\left( t, a^3 \langle C \rangle, D / a^2, \dot \gamma,
        \left\{ l_i / a \right\} \right)
\end{eqnarray}
holds. The ``Hamiltonian'' of the previous subsection is of course
subject to the same scale invariance. This means that for each point
in parameter space there is a whole ``iso-line'' in parameter space
that fits the data just as well as the original point. Therefore, in
order to improve the MC sampling, we generated such an iso-line for
each point in parameter space that was produced by the Markov chain of
the previous subsection. Of course, the iso-lines were confined to the
region of the overall parameter box. It turned out that our Markov
chains were still so short that this improvement was not completely
superfluous (as it would be in the limit of very long chains). In
other words, taking the invariance into account helped us to avoid
underestimating the errors.

In practice, this was done as follows: Assuming that the most accurate
input parameters are the penetration depth $d_p = 100 \pm 10 nm$, the
diffusion constant $D=36 \pm 5 \mu m^2/s$ and the separation distance
$s_x = 800 \pm 80 nm$, we calculate for every data point a minimum and
a maximum scaling factor $a$, such that we obtain $d_p^{(min)} <
a^{-1} d_p < d_p^{(max)}$, $s_x^{(min)} < a^{-1} s_x < s_x^{(max)}$
and $D^{(min)} < a^{-2} D < D^{(max)}$, for all $a$ in
$(a_{min},a_{max})$. This provides us with additional data points in
parameter space that are added to the statistics.

\subsection{Sample-to-Sample Fluctuations}

\begin{table}[t]
  \noindent
  \begin{centering}
    \begin{tabular}{|l||c c|c c|c c|}
      \hline
      &&&&&&
      \tabularnewline
      seed & 42 && 4711 && 2409 &
      \tabularnewline
      \hline
      & av & $\sigma$ & av & $\sigma$ & av & $\sigma$
      \tabularnewline
      \hline
      &&&&&&
      \tabularnewline
      $\langle C \rangle_A [\mu m^{-3}]$ & $17.83$ & $1.92$ & $17.81$ & $1.96$ &
      $17.94$ & $1.94$
      \tabularnewline
      $\langle C \rangle_C [\mu m^{-3}]$ & $16.53$ & $1.80$ & $16.45$ & $1.82$ &
      $16.83$ & $1.83$
      \tabularnewline
      $d_p [nm]$ & $95.81$ & $3.50$ & $96.02$ & $3.58$ & $95.78$ & $3.52$
      \tabularnewline
      $(k NA)^{-1} [nm]$ & $68.70$ & $2.60$ & $68.58$ & $2.61$ & $69.81$ & $2.69$
      \tabularnewline
      $s_x [nm]$ & $781.94$ & $27.70$ & $779.54$ & $28.28$ & $793.22$ & $28.24$
      \tabularnewline
      $D [\mu m^2/s]$ & $36.59$ & $2.56$ & $36.47$ & $2.62$ & $36.63$ & $2.57$
      \tabularnewline
      $l_s + R[nm]$ & $12.80$ & $1.10$ & $11.62$ & $0.90$ & $15.14$ & $1.21$
      \tabularnewline
      $\xi$ & $1.441$ & $0.018$ & $1.277$ & $0.016$ & $1.718$ & $0.017$
      \tabularnewline
      acceptance rate & $82.0 \%$ && $82.0 \%$ && $82.2 \%$ &
      \tabularnewline
      No. of MC steps & $609410$ && $609590$ && $610510$ &
      \tabularnewline
      \hline
    \end{tabular}
    \par
  \end{centering}
  \caption{Averaged values (av) and standard deviations ($\sigma$) calculated
    from MC simulations with fixed $\dot \gamma = 3800 s^{-1}$, but different
    start values (``seeds'') for the random number generator.} 
  \label{tab:influence_seed}
\end{table}

It should be noted that the parameters found by the procedure outlined
above are optimized for a specific set of random numbers used to
generate the trajectories. Therefore, one must expect that one obtains
different results when changing the set of random numbers. In our
statistical-mechanical picture, we may view the set of random numbers
as random ``coupling constants'' of a disordered system like a spin
glass \cite{mezard}, where the disorder is weak since the number of
trajectories is large. For disordered systems, the phenomenon of
``sample-to-sample'' fluctuations is well-known, and it should be
taken into account. We have therefore run one test where we applied
the same analysis to three different random number sequences. Indeed
we found (see Tab.~\ref{tab:influence_seed}) that sample-to-sample
fluctuations are observable, and somewhat larger than the errors
obtained from simple MC, while still being of the same order of
magnitude. A conservative error estimate should therefore take these
fluctuations into account, by multiplying the error estimates from
plain MC by, say, a factor of three. In what follows, we will only
report the simple MC estimates for the errors.

\section{Results}
\label{sec:results}
     
\begin{figure}[t]
   \noindent
   \begin{centering}\hspace{-0.4cm}
      \includegraphics[clip,width=0.49\textwidth]
      {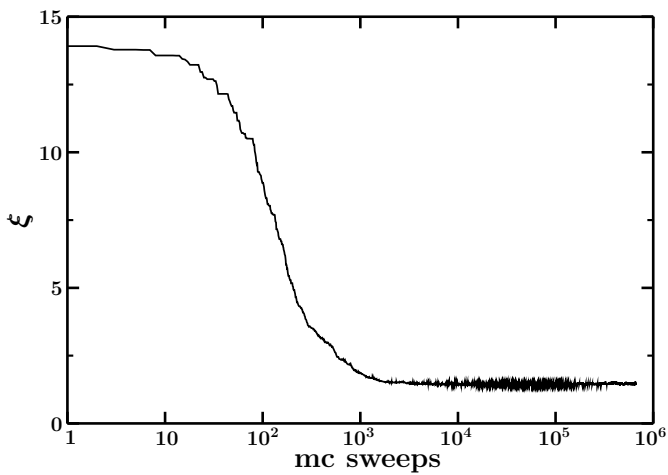}
   \end{centering}
   \caption{Goodness of simulation $\xi$ as function of the number of
      Monte Carlo steps for $\dot \gamma = 3800 s^{-1}$.}
   \label{fig: mc xi vs sweep}
\end{figure}

\begin{figure}[t]
   \noindent
   \begin{centering}\hspace{-0.4cm}
     \includegraphics[clip,width=0.49\textwidth]
     {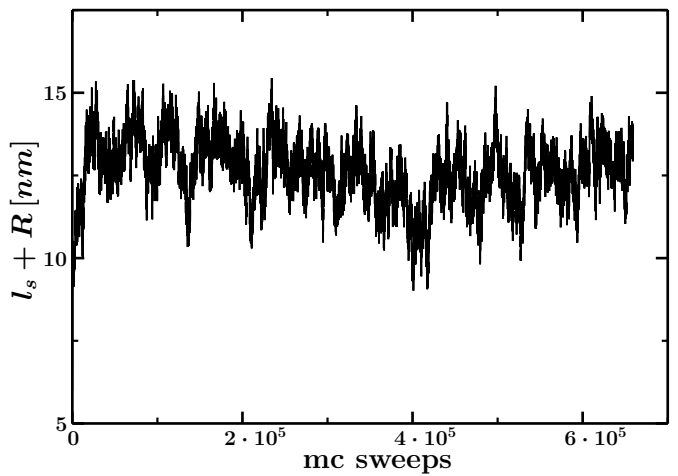}
   \end{centering}
   \caption{Slip length plus particle radius as function of the number of Monte
     Carlo steps for $\dot \gamma = 3800 s^{-1}$.}
   \label{fig: mc slip vs sweep}
\end{figure}

The experiments were performed with a penetration depth of the
evanescent wave of $d_p \simeq 100nm$, the lateral size of the
observation volumes (within the Gaussian model) was $w_0 \simeq 250nm$
and their center-to-center separation was $s_x \simeq 800nm$. 
Furthermore, the diffusion constant of the tracers is known to be
roughly $D \simeq 36 \mu m^2 / s$ as measured by dynamic light
scattering. The shear rate was determined from an independent
measurement using single-focus confocal FCS \cite{Dit02, Kohler2000}
where the entire flow profile across the microchannel was mapped
out. Alternatively, one might also use double-focus confocal FCCS
\cite{Bri99, Lum03}. From this measurement, we obtained a shear rate
at the bottom channel wall of $\dot \gamma = 3854 \pm 32 s^{-1}$. More
details on this issue and some theoretical background are presented in
the appendix. Nevertheless, we took a conservative approach and
allowed the shear rate to vary between $3500 s^{-1}$ and $4000
s^{-1}$. Finally, we expected the slip length to be not more than a
few nanometers, but we nevertheless allowed it to vary up to $\simeq
100 nm$. These estimates allowed us to start the Monte Carlo procedure
with good input values.

We then observed the Monte Carlo simulation to systematically drift to
smaller and smaller values of $\dot \gamma$, until finally ``getting
stuck'' at the imposed lower boundary, $\dot \gamma = 3500 s^{-1}$.
What we mean by this term is a behavior where fluctuations near $3500
s^{-1}$ still occur, but in such a way that $3500 s^{-1}$ is the most
probable value, while smaller values only do not occur because we do
not allow them. Since we know experimentally that $\dot \gamma = 3500
s^{-1}$ is clearly unacceptable, this behavior again indicates that
the theoretical model is not completely sufficient to describe the
experimental data (see also the discussion in
Secs.~\ref{sec:exp_corrfunc} and \ref{sec:subtraction}).

We therefore decided to keep $\dot \gamma$ fixed during a Monte Carlo
run, and rather vary it systematically in the given range. For none of
the parameters were we able to obtain a better goodness of simulation
than $\xi \simeq 1.25$, which is still a bit too large, i.~e.
indicates a non-perfect fit (although the data on $(G_{12} +
G_{21})/2$ have been discarded already). The convergence behavior of
the method is shown in Fig. \ref{fig: mc xi vs sweep}, where we plot
$\xi$ as a function of the number of Monte Carlo iterations. For the
Gaussian model, the best $\xi$ value that we could obtain was $\xi
\simeq 2.5$, which is substantially worse.

\begin{figure}[t]
   \noindent
   \begin{centering}\hspace{-0.4cm}
      \includegraphics[clip,width=0.49\textwidth]
      {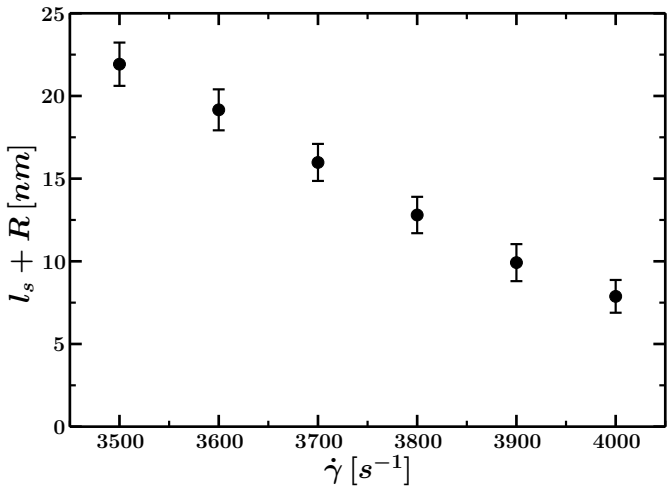}
   \end{centering}
   \caption{Averaged slip length as function of the shear rate,
     calculated from the Monte Carlo results.}
   \label{fig: mc slip vs shear}
\end{figure}

With these caveats in mind, we may proceed to study the parameter
values that the Monte Carlo procedure yields. Obviously, the most
interesting one is the slip length $l_s$, or the sum $l_s + R$ (recall
that the method does not provide an independent estimate for these
parameters, but only for their sum). Figure \ref{fig: mc slip vs
  sweep} presents data on the evolution of $l_s + R$ during the Monte
Carlo process for $\dot \gamma = 3800 s^{-1}$; $l_s + R$ is thus seen
to fluctuate between roughly $10 nm$ and $15 nm$, which is,
\emph{within the limitations of the model}, the statistical
experimental uncertainty of this quantity. The mean and standard
deviation of $l_s + R$ is shown in Fig. \ref{fig: mc slip vs shear} as
a function of shear rate, which are thus clearly seen to not be
independent. Since we know $\dot \gamma$ much more accurately than the
range plotted in Fig. \ref{fig: mc slip vs shear}, we see that in
principle a fairly accurate determination of $l_s$ is possible, if the
underlying theoretical model is detailed enough to fully describe the
physics. One should note that the particle size $R$ (more precisely,
its hydrodynamic radius) is roughly $7 nm$; taking this into accunt as
well, we find a value that is clearly smaller than $10 nm$. One should
also note that for the Gaussian model we obtained a very similar
curve; however, here the $l_s + R$ values are systematically smaller
by roughly $5 nm$. This again highlights the importance of having an
accurate model for the MDE.

The other results obtained from our MC fits are reported in
Tab. \ref{tab:MC_results_PCPSF}.

\begin{table}
  \noindent
  \begin{centering}
    \begin{tabular}{|l||c c|c c|c c|}
      \hline
      &&&&&&
      \tabularnewline
      $\dot \gamma [s^{-1}]$ & 3500 && 3600 && 3700 &
      \tabularnewline
      \hline
      & av & $\sigma$ & av & $\sigma$ & av & $\sigma$
      \tabularnewline
      \hline
      &&&&&&
      \tabularnewline
      $\langle C \rangle_A [\mu m^{-3}]$ & $17.82$ & $1.90$ & $17.91$ & $1.88$ &
      $17.93$ & $1.89$
      \tabularnewline
      $\langle C \rangle_C [\mu m^{-3}]$ & $19.94$ & $1.82$ & $17.00$ & $1.80$ &
      $16.82$ & $1.78$
      \tabularnewline
      $d_p [nm]$ & $95.83$ & $3.46$ & $95.66$ & $3.42$ & $95.64$ & $3.44$
      \tabularnewline
      $(k NA)^{-1} [nm]$ & $68.84$ & $2.53$ & $69.10$ & $2.55$ & $69.01$ & $2.56$
      \tabularnewline
      $s_x [nm]$ & $774.39$ & $26.59$ & $777.98$ & $26.63$ & $780.53$ & $26.97$
      \tabularnewline
      $D [\mu m^2/s]$ & $36.71$ & $2.49$ & $36.76$ & $2.48$ & $36.72$ & $2.51$
      \tabularnewline
      $l_s + R[nm]$ & $21.92$ & $1.31$ & $19.16$ & $1.24$ & $15.98$ & $1.12$
      \tabularnewline
      $\xi$ & $1.377$ & $0.016$ & $1.40$ & $0.016$ & $1.420$ & $0.017$
      \tabularnewline
      acceptance rate & $82.1 \%$ && $82.1 \%$ && $82.1 \%$ &
      \tabularnewline
      No. of MC steps & $608090$ && $611290$ && $610160$ &
      \tabularnewline
      \hline
      \hline
      &&&&&&
      \tabularnewline
      $\dot \gamma [s^{-1}]$ & 3800 && 3900 && 4000 &
      \tabularnewline
      \hline
      & av & $\sigma$ & av & $\sigma$ & av & $\sigma$
      \tabularnewline
      \hline
      &&&&&&
      \tabularnewline
      $\langle C \rangle_A [\mu m^{-3}]$ & $17.83$ & $1.92$ & $17.67$ & $2.00$ &
      $17.96$ & $1.89$
      \tabularnewline
      $\langle C \rangle_C [\mu m^{-3}]$ & $16.53$ & $1.80$ & $16.21$ & $1.86$ &
      $16.41$ & $1.74$
      \tabularnewline
      $d_p [nm]$ & $95.81$ & $3.50$ & $96.14$ & $3.70$ & $95.58$ & $3.41$
      \tabularnewline
      $(k NA)^{-1} [nm]$ & $68.70$ & $2.60$ & $68.42$ & $2.71$ & $69.01$ & $2.56$
      \tabularnewline
      $s_x [nm]$ & $781.94$ & $27.70$ & $783.75$ & $28.76$ & $789.77$ & $27.35$
      \tabularnewline
      $D [\mu m^2/s]$ & $36.59$ & $2.56$ & $36.46$ & $2.64$ & $36.71$ & $2.51$
      \tabularnewline
      $l_s + R [nm]$ & $12.80$ & $1.10$ & $9.92$ & $1.12$ & $7.88$ & $0.99$
      \tabularnewline
      $\xi$ & $1.441$ & $0.018$ & $1.464$ & $0.019$ & $1.477$ & $0.016$
      \tabularnewline
      acceptance rate & $82.0 \%$ && $81.9 \%$ && $82.0 \%$ &
      \tabularnewline
      No. of MC steps & $609410$ && $610330$ && $610440$ &
      \tabularnewline
      \hline
    \end{tabular}
    \par
  \end{centering}
  \caption{Averaged values (av) and standard deviations ($\sigma$) calculated
	   from MC simulations with various shear rates.} 
  \label{tab:MC_results_PCPSF}
\end{table}

Clearly, the $l_s$ values of Fig. \ref{fig: mc slip vs shear} could
only be viewed as definitive experimental results on $l_s$ if the
agreement between experiment and model were perfect, with $\xi \simeq
1$, and a good fit of \emph{all} correlation functions. The reasons
for the observed deviations are not completely clear; however, all our
findings hint very strongly to deficiencies in the description of the
observation volumes, i.~e. too inaccurate modeling of the detailed
optical phenomena that finally give rise to the shape of these
functions. Nevertheless, one should also bear in mind that the dynamic
model is also rather simple, neglecting both hydrodynamic and residual
electrostatic interactions with the wall. While one must expect that
further refinements of the model will change both the $l_s$ values as
well as their error bars, we believe that it is not probable that such
a change would be huge. Given all the various systematic uncertainties
of the modeling, we would, in view of our data, not exclude a
vanishing slip length, while we consider a value substantially larger
than, say, $15 nm$ as fairly unlikely.

Let us conclude this section by a few more remarks concerning our
choice of parameters and the systematic errors of the method. From the
setup it is clear that there are three parameters that can be varied
experimentally fairly easily --- these are the shear rate $\dot
\gamma$, the penetration depth $d_p$, and the effective
pinhole-pinhole distance in sample space, $s_x$. The choice of
parameters was governed by various experimental considerations, which
we will attempt to explain in what follows.

It is clear that one wants a fairly large shear rate $\dot \gamma$, in
order to ensure that the signal has a sizeable contribution from flow
effects. In practice, however, increasing $\dot \gamma$ further by a
substantial amount is limited by experimental constraints, such as
channel construction, beaker elevation, etc. Furthermore, the choice of
$d_p$ is subject to similarly severe experimental constraints:
Increasing $d_p$ substantially would mean that we would approach the
limit angle of total reflection closely, which would result in a very
inaccurate a priori estimate of $d_p$. On the other hand, an even
smaller penetration depth value would be too close to the limits of
the capabilities of the objective, resulting in possible optical
distortion effects which we would like to avoid. Finally, the choice
of $s_x$ was governed by our early attempts to suppress overlap
effects by simply picking a fairly large value, such that the overlap
integral is small. There are however two problems about such an
idea. Firstly, a large value of $s_x$ decreases not only the overlap,
but also the cross-correlation function as a whole \cite{Koy09}, such
that it ultimately becomes impossible to sample the data with
sufficient statistical accuracy on the time scale on which we can
confidently keep the experimental conditions stable. Our value of
$s_x$ should therefore be viewed as limited by such
considerations. However, secondly, and more importantly, we realized
in the course of our analysis that the overlap issue is \emph{not} a
problem of an unintelligent choice of parameters at all, but rather of
our insufficient theoretical modeling of the MDE functions. As we
have seen above, our results for the slip length depend rather
sensitively and fairly substantially on the choice of the MDE function
(up to nearly a factor of two). In our opinion, there is no reason to
assume that this dependence would go away if we had picked parameters
in a regime of small or vanishing overlap. From this perspective, we
view the overlap essentially as a blessing, since it shows us where
the main source of systematic error is probably located.

In the light of these remarks, it would of course be interesting to
systematically investigate the influence of the parameters $d_p$ and
$s_x$ on our results. In terms of the correlation functions as such,
this has been done in Ref. \cite{Koy09}, and we refer the interested
reader to that paper. However, doing the full analysis for a whole
host of parameters would imply a very substantial amount of work,
since all the experimental curves would have to be re-sampled again,
in order to meet the rather stringent requirement of statistical
accuracy that is built into our approach. We have hence not attempted
to do this, but rather believe that it will be more fruitful to
concentrate the efforts of future work on attempts to improve the
theoretical MDE modelling, even if that will be challenging. As far as
the slip length is concerned, one must of course expect that the
fitted value will depend on parameters such as $d_p$ and $s_x$, but
only to the extent that this reflects the systematic error --- if the
physics were modeled perfectly correctly, we would of course always
obtain the same value.

\section{Conclusions}
\label{sec:conclus}
    
The results from the previous sections demonstrate that the method of
TIR-FCCS in combination with the presented Brownian Dynamics and Monte
Carlo based data analysis is in principle a very powerful tool for the
analysis of hydrodynamic effects near solid-liquid interfaces. Already
within the investigated simple model of the present paper, we can
conclude that the slip length at our hydrophilic surface is not more
than $10nm$. It was only the data processing via the Brownian Dynamics
/ Monte Carlo analysis that was able to demonstrate how highly
sensitive and accurate TIR-FCCS is.

The computational method has the advantage to be easily extensible to
include more complex effects. For example, the hydrodynamic
interactions of the particles with the wall would cause an anisotropy
in the diffusion tensor \cite{Dhont07} and a $z$ dependence,
electrostatic interactions would give rise to an additional force term
in the Langevin equation, while polydispersity could be investigated
by randomizing the particle size and the diffusion properties
according to a given distribution. While these contributions are
expected to yield a further improvement of the method, this was not
attempted here, and is rather left to future investigations. However,
we have also identified the inaccuracies in modeling the observation
volumes as (most probably) \emph{the} main bottleneck in finding good
agreement between theory and experiment, i.~e., at present, as the
main source of systematic errors, which makes it difficult to find a
fully reliable error bound on the value of the slip length.

Conversely, the problem of dealing with statistical errors can be
considered as solved. For an extensive data analysis, as it has been
presented here, one may need a supercomputer in order to obtain highly
accurate results in fairly short time. Nevertheless, the method will
yield meaningful results even if confined to just a single modern
desktop computer. Given the moderate amount of computer time on a
high-performance machine, one should expect that quite accurate data
should be obtainable within reasonable times by making use of the
powerful newly emerging GPGPU cards.

In our opinion, the presented method is a conceptually simple and
widely applicable approach to process TIR-FCCS data, that is clearly
only limited by inaccurate modeling. We believe that it has the
potential to become the standard and general tool to process such
data, in particular as soon as the optics is understood in better
detail. The principle as such is applicable to all kinds of
correlation techniques, such as FCS/TIR-FCS etc., and we think it is
the method of choice whenever one investigates a system whose
complexity is beyond analytical treatment.

\acknowledgments{This work was funded by the SFB TR 6 of the
  Deutsche Forschungsgemeinschaft. Computer time was provided
  by Rechenzentrum Garching. We thank J. Ravi Prakash
  and A. J. C. Ladd for helpful discussions.}

\begin{appendix}

\section{Solution of the Stokes Equation
in a Rectangular Channel}
    
The flow profile throughout the height of the microchannel was
measured by single-focus FCS under the same conditions as the TIR-FCCS
experiments; the result is shown in Fig. \ref{fig: hphil flow
  profile}. From a fit via a Poiseuille profile (solid line), we
obtained an independent estimator for the shear rate near the
wall, $\dot \gamma = 3854 \pm 32 s^{-1}$.

\begin{figure}[t]
   \noindent
   \begin{centering}\hspace{-0.4cm}
      \includegraphics[clip,width=0.49\textwidth]
      {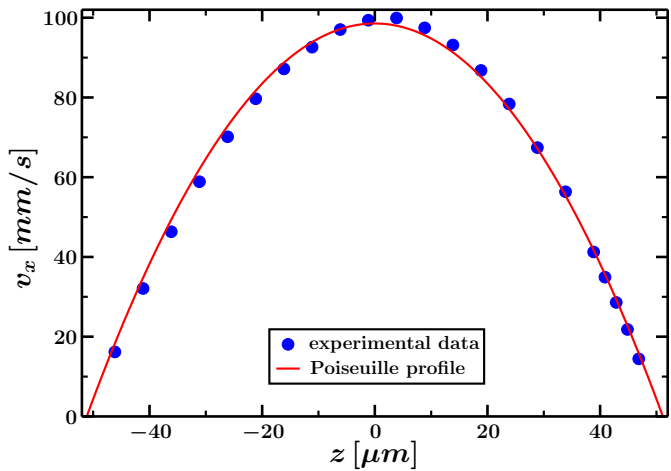}
   \end{centering}
   \caption{(Color online) Flow profile and Poiseuille fit along
     $z$-direction (surface of measurement is located at $z\simeq 50
     \mu m$).}
   \label{fig: hphil flow profile}
\end{figure}

The purpose of this appendix is to analyze the theoretical background
of this fit in some more detail. For a pure Poiseuille flow, i.~e. a
simple parabolic profile, it is clear that the shear rate at the
surface does not depend on the slip length $l_s$, because in this case
a finite $l_s$ value simply shifts the profile by a constant
amount. Therefore, in this case $l_s$ is indeed irrelevant for the
fit. A short discussion on such issues is also found in Ref.
\cite{Lum03}, and experimentally \cite{Jos05,Vin09} it is also known
that typically the shift is so small that a finite slip length is
hard to detect by direct measurements of the profile. However, from a
theoretical and quantitative point of view it is not quite clear how
well it is justified to assume a strictly parabolic profile, i.~e. to
assume that the flow extends infinitely in $y$ direction --- in our
experiments, $L_y / L_z = 40$, which is large but not infinite.
For finite values of $L_y / L_z$, the profile is somewhat distorted,
and if this distortion is sufficiently large, then also a possible
effect of $l_s$ should be taken into account. These questions can be
easily answered by solving the flow problem in a rectangular channel
in the presence of slip exactly, and this shall be done in what
follows. The result of this analysis will be that for our conditions
the distortion of the profile is indeed completely negligible, and
that therefore $l_s$ needs not be taken into account either.

We start by considering the Stokes equation
\begin{equation}\label{eq: 2d Stokes}
  \eta \left( \frac{\partial^2}{\partial y^2}
  + \frac{\partial^2}{\partial z^2} \right) v_x (y,z) + f = 0 ,
\end{equation}
in a rectangular channel with dimensions $[- L_y / 2, L_y / 2]
\times [- L_z / 2, L_z / 2]$ in the $yz$-plane, as in the
experiment. Here, $\eta$ is the viscosity of the liquid and $f$ is the
driving force density or pressure gradient acting on the liquid in
$x$-direction. We assume that all surfaces have the same slip length.

For the case of a no-slip boundary condition, the solution has been
given in the textbook of Spurk and Aksel \cite{Spu06}, however in a
form that does not explicitly spell out the symmetry under exchange of
$y$ and $z$. Here we give the solution in a form that shows that
symmetry, and generalize it to the case of a nonvanishing slip length
$l_s$.

Using the methods and notation of quantum mechanics, and allowing for
some minor amount of numerics to evaluate a series, the solution is
simple and straightforward. We identify a function $f(y,z)$ with a
vector $\left\vert f \right>$ in a Hilbert space, and define the
scalar product as
\begin{equation}
\left< f \vert g \right> =
\int_{-L_y/2}^{+L_y/2} dy \int_{-L_z/2}^{+L_z/2} dz f^\star (y,z) g(y,z) .
\end{equation}
Defining a ``Hamilton operator'' via
\begin{equation}
{\cal H} = - \frac{\eta}{f}
  \left( \frac{\partial^2}{\partial y^2}
  + \frac{\partial^2}{\partial z^2} \right) ,
\end{equation}
the Stokes equation is written as
\begin{equation}
{\cal H} \left\vert v_x \right> = \left\vert 1 \right> .
\end{equation}
Obviously, the functions
\begin{equation}
\left\vert k_y, k_z \right> = N(k_y,k_z) \cos (k_y y) \cos (k_z z)
\end{equation}
with $k_y > 0$, $k_z > 0$ and
\begin{eqnarray} \label{eq: eigenfunction normalization}
\nonumber
N(k_y,k_z) & = &
\left[ \frac{L_y}{2} + \frac{\sin (k_y L_y)}{2 k_y} \right]^{-1/2} \\
& \times &
\left[ \frac{L_z}{2} + \frac{\sin (k_z L_z)}{2 k_z} \right]^{-1/2}
\end{eqnarray}
are normalized eigenfunctions of ${\cal H}$,
\begin{equation}
{\cal H} \left\vert k_y, k_z \right> =
\frac{\eta}{f} \left( k_y^2 + k_z^2 \right)
\left\vert k_y, k_z \right>
\end{equation}
with
\begin{equation}
\left< k_y, k_z \vert q_y, q_z \right> = \delta_{k_y,q_y} \delta_{k_z,q_z} .
\end{equation}

\begin{figure}[t]
   \noindent
   \begin{centering}
     \hspace{-0.4cm}
     \includegraphics[clip,width=0.49\textwidth]
     {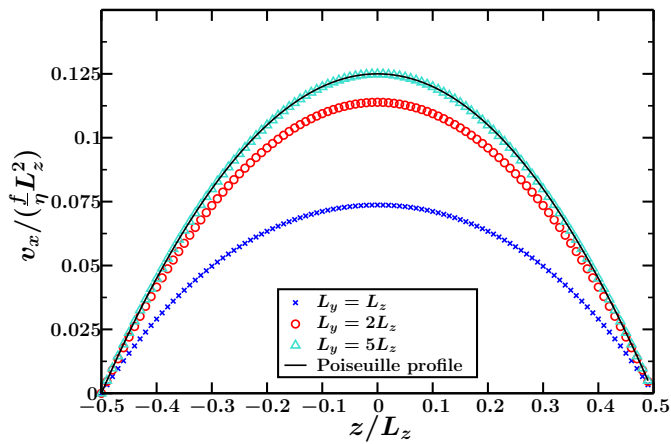}
   \end{centering}
   \caption{(Color online) One-dimensional cut of the flow profile at
     $y=0$ for no-slip boundary conditions and several values of
     $L_y$.}
   \label{fig: flow profile Ly series}
\end{figure}

\begin{figure}[t]
   \noindent
   \begin{centering}
      \hspace{-0.4cm}
      \includegraphics[clip,width=0.49\textwidth]
      {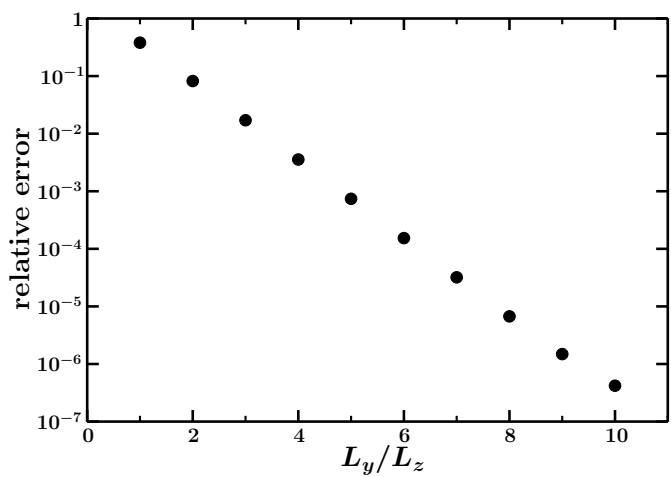}
   \end{centering}
   \caption{Averaged deviation between a one-dimensional cut of the
     flow profile at $y=0$ for no-slip boundary conditions and the
     Poiseuille solution, as function of the width-height ratio of the
     channel.}
   \label{fig: flow profile Ly error}
\end{figure}

The eigenfunctions must satisfy the boundary conditions, and hence
the discrete wave numbers $k_y$, $k_z$ must be the solutions of the
transcendental equations
\begin{subequations}
  \label{eq: transcendental boundary condition}
  \begin{eqnarray}
    l_s k_y & = & \cot \left( k_y \frac{L_y}{2} \right) , \\
    l_s k_z & = & \cot \left( k_z \frac{L_z}{2} \right) ,
  \end{eqnarray}
\end{subequations}
which, in the general case, can be found numerically. In the no-slip
case, the solutions are simply given by $k_y = \pi / L_y, 3 \pi / L_y,
\ldots$ and analogously for $k_z$. Equation \ref{eq: transcendental
  boundary condition} allows us to re-write Eq.  \ref{eq:
  eigenfunction normalization} as
\begin{eqnarray}
\nonumber
N(k_y,k_z) & = &
\left[ \frac{L_y}{2} 
+ l_s \sin^2 \left( k_y \frac{L_y}{2} \right) \right]^{-1/2} \\
& \times &
\left[ \frac{L_z}{2} 
+ l_s \sin^2 \left( k_z \frac{L_z}{2} \right) \right]^{-1/2} .
\end{eqnarray}
Since the set of eigenfunctions is complete, the spectral
representations of ${\cal H}$ and ${\cal H}^{-1}$ are given by
\begin{eqnarray}
  {\cal H} & = & \frac{\eta}{f} \sum_{k_y, k_z} 
  \left( k_y^2 + k_z^2 \right)
  \left\vert k_y, k_z \right> \left< k_y, k_z \right\vert , \\
  {\cal H}^{-1} & = & \frac{f}{\eta} \sum_{k_y, k_z} 
  \left( k_y^2 + k_z^2 \right)^{-1}
  \left\vert k_y, k_z \right> \left< k_y, k_z \right\vert ,
\end{eqnarray}
resulting in the solution
\begin{eqnarray}
  \left\vert v_x \right> 
  & = & 
  {\cal H}^{-1} \left\vert 1 \right> \\
  \nonumber
  & = &
  \frac{f}{\eta} \sum_{k_y, k_z} \left( k_y^2 + k_z^2 \right)^{-1}
  \left< k_y, k_z \vert 1 \right> \left\vert k_y, k_z \right> \\
  \nonumber
  & = &
  \frac{f}{\eta} \sum_{k_y, k_z} \left( k_y^2 + k_z^2 \right)^{-1}
  N(k_y,k_z)^2 \frac{4}{k_y k_z} \\
  \nonumber
  & \times &
  \sin \left( k_y \frac{L_y}{2} \right)
  \sin \left( k_z \frac{L_z}{2} \right)
  \cos (k_y y) \cos (k_z z) .
\end{eqnarray}

Figure \ref{fig: flow profile Ly series} shows the resulting flow
profile at $y = 0$ (in the center of the channel) as a function of
$z$, for vanishing slip length and various width-to-height ratios
$L_y/L_z$ of the channel. One sees that the convergence to the
asymptotic Poiseuille profile $v_{\mathrm{P}}$ is indeed extremely
rapid. The deviation, defined via
\begin{equation}
   \mathrm{relative\;error}
   = \frac{1}{L_z} \int_{-L_z/2}^{L_z/2} dz
   \left\vert
   \frac{v_x (0,z) - v_{\mathrm{P}} (z)}{v_{\mathrm{P}} (z)}
   \right\vert ,
\end{equation}
is displayed as a function of $L_y/L_z$ in Fig. \ref{fig: flow profile
  Ly error}. The rate of convergence is apparently exponential, and
for the experimental value $L_y / L_z = 40$ the deviation is
seen to be much smaller than the resolution of the measurements.
Therefore, the assumption of a parabolic profile is indeed justified.

\end{appendix}


\end{document}